\documentclass[twocolumn,showpacs,amsfonts,amssymb,amsmath,floats,superscriptaddress,aps]{revtex4-1}

\usepackage{graphicx}
\usepackage{upgreek}

\usepackage{hyperref}
\usepackage{todonotes}
\usepackage[normalem]{ulem}

\def\bra#1{\left.\left\langle#1\right.\right|}
\def\ket#1{\left.\left|#1\right.\right\rangle}
\def\braket#1#2#3{\left\langle#1\middle|#2\middle|#3\right\rangle}
\def\up{\uparrow}
\def\down{\downarrow}

\def\e{\varepsilon}

\def\im{{\rm i}}
\def\muB{\mu_{B}}
\def\ge{g_{e}}
\def\muN{\mu_{N}}
\def\gN{g_{N}}
\def\Hcsm{H_{\rm CSM}}
\def\Bext{B_{\rm ext}}
\def\bnx{B_{N}^x}
\def\bnxj{B_{{N},j}^x}
\def\TR{T_{R}}
\def\NP{N_{P}}
\def\NC{N_{C}}

\def\expect#1{\left\langle#1\right\rangle}

\def\Sz{\expect{S^z}}
\def\Szt{\expect{S^z (t)}}
\def\SzTR{\expect{S^z \left( N_{P} T_{R} \right)}}
\def\SyTR{\expect{S^y \left( N_{P} T_{R} \right)}}
\def\SpTR{S^\perp \left( N_{P} T_{R} \right)}
\def\Tr#1{{\rm Tr}\left[#1\right]}

\begin{document}

\title{Magnetic field dependence of the electron spin revival amplitude in periodically pulsed quantum dots}

\author{Iris Kleinjohann}
\address{Lehrstuhl f\"ur Theoretische Physik II, Technische Universit\"at Dortmund,
Otto-Hahn-Stra{\ss}e 4, 44227 Dortmund, Germany}

\author{Eiko Evers}
\address{Experimentelle Physik IIa, Technische Universit\"at Dortmund,
Otto-Hahn-Stra{\ss}e 4a, 44227 Dortmund, Germany}

\author{Philipp Schering}
\address{Theoretische Physik I, Technische Universit\"at Dortmund,
Otto-Hahn-Stra{\ss}e 4, 44227 Dortmund, Germany}

\author{Alex Greilich}
\address{Experimentelle Physik IIa, Technische Universit\"at Dortmund,
Otto-Hahn-Stra{\ss}e 4a, 44227 Dortmund, Germany}

\author{G\"otz S.\ Uhrig}
\address{Theoretische Physik I, Technische Universit\"at Dortmund,
Otto-Hahn-Stra{\ss}e 4, 44227 Dortmund, Germany}

\author{Manfred Bayer}
\address{Experimentelle Physik IIa, Technische Universit\"at Dortmund,
Otto-Hahn-Stra{\ss}e 4a, 44227 Dortmund, Germany}
\author{Frithjof B.\ Anders}
\address{Lehrstuhl f\"ur Theoretische Physik II, Technische Universit\"at Dortmund,
Otto-Hahn-Stra{\ss}e 4, 44227 Dortmund, Germany}

\date{\today}

\begin{abstract}

Periodic laser pulsing of singly charged semiconductor quantum dots in an external magnetic field leads to a synchronization of the spin dynamics with the optical excitation.
The pumped electron spins partially rephase prior to each laser pulse, causing a revival of electron spin polarization with its maximum at the incidence time of a laser pulse.
The amplitude of this revival is amplified by the frequency focusing of the surrounding nuclear spins.
Two complementary theoretical approaches for simulating up to 20 million laser pulses are developed and employed that are able to bridge between 11 orders of magnitude in time:
a fully quantum mechanical description limited to small nuclear bath sizes and a technique based on the classical equations of motion applicable for a large number of nuclear spins.
We present experimental data of the nonmonotonic revival amplitude as function of the magnetic field applied perpendicular to the optical axis.
The dependence of the revival amplitude on the external field with a profound minimum at $4\;$T is reproduced by both of our theoretical approaches and is ascribed to the nuclear Zeeman effect.
Since the nuclear Larmor precession determines the electronic resonance condition, it also defines the number of electron spin revolutions between pump pulses, the orientation of the electron spin at the incidence time of a pump pulse, and the resulting revival amplitude.
The magnetic field of $4\;$T, for example, corresponds to half a revolution of nuclear spins between two laser pulses.

\end{abstract}

\maketitle

\section{Introduction}

Manipulation of the resident electron spins in singly charged semiconductor quantum dots (QDs) using laser pulses is considered a promising route for optically controlled quantum functionality \cite{PhysRevLett.96.227401,*GreilichScience2007}.
The well localized electron spins exhibit an increased coherence time, which is primarily limited by the hyperfine interaction between the electron spin and the surrounding nuclear spins at cryogenic temperatures \cite{MerkulovEfrosRosen,CoishLoss2004,FischerLoss2008,HansonSpinQdotsRMP2007,Dyakonov,Urbaszek2013}. Periodic optical pumping in an external magnetic field leads to a synchronization of the electron spin precession frequencies to the pumping periodicity by nuclear frequency focusing.
The Floquet's theorem predicts resonance or mode-locking conditions \cite{PhysRevLett.96.227401} that have been investigated using a classical representation of the spin dynamics \cite{PetrovYakovlev2012,Jaeschke2017} as well as a perturbative quantum mechanical treatment of the spin system \cite{BeugelingUhrigAnders2016,*BeugelingUhrigAnders2017}.
At resonance, the electron spins partially rephase prior to each laser pulse causing a constructive interference.
Since each electron is well localized within its own bath of nuclear spins, the electron spin and the nuclear spins evolve as a coupled system reaching a stroboscopic stationary state after long pumping \cite{PhysRevLett.96.227401}.
This quasi-stationary state of a periodically pumped ensemble of QDs strongly differs from the equilibrium starting point and is characterized by the synchronization of the evolution of electronic and nuclear spins implying a finite revival amplitude of the electron spin polarization.

Although the electronic resonance condition in steady-state is well established \cite{PhysRevLett.96.227401,PetrovYakovlev2012,Jaeschke2017,BeugelingUhrigAnders2017}, the dependency of the revival amplitude on the applied magnetic field has not been thoroughly investigated yet.
In this paper, we approach the subject in a threefold way.
After briefly presenting recent experimental measurements of the revival amplitude, we devise a full quantum mechanical approach to the numerical calculation of a periodically pulsed QD.
The results of the quantum mechanical exploration are supplemented by a classical approach \cite{Schering2018}.

The theoretical approaches have to face the challenge of a wide variation of time scales in the pulsed QD system.
Short laser pulses with a duration of two to ten picoseconds have to be combined with free dynamics of $13.2 \,$ns between the laser pulses to a repetitive propagation in time.
Our approaches achieve the simulation of up to $20$ million laser pulses, hence, covering a total simulation time up to $0.2 \,$s and bridging between eleven orders in magnitude.
This huge computational effort is necessary to reach a converged steady-state of the spin dynamics, which is crucial to analyze the revival amplitude and its dependence on the external magnetic field.

Both theoretical treatments base on the central spin model (CSM) \cite{Gaudin} containing the hyperfine interaction between the resident electron spin and the surrounding nuclear spins as well as the Zeeman effect.
The quantum mechanical approach includes the full quantum mechanical time evolution of the density operator and hence focuses on a rather small nuclear bath of $N=6$ nuclear spins.
However, it has been established \cite{alhas06,HackmannAnders2014,FroehlingAnders2017,BeugelingUhrigAnders2016,BeugelingUhrigAnders2017} that even for a low number of nuclei the generic spin dynamics \cite{MerkulovEfrosRosen} can be reproduced.
The time evolution between laser pulses is captured by the exact solution of a Lindblad equation accounting for the decay of the optically excited trion and the dynamics of the CSM.

The laser pulses are quantum mechanically described by unitary transformations.
For this purpose, we first treat the laser pulse in the limit of vanishing duration considering the pulses as instantaneous.
However, a main advantage of our quantum mechanical approach is the possibility to lift this approximation and turn towards pulses with arbitrary duration and shape.
In the later part of the paper, Gaussian pump pulses with a width of several picoseconds, which are based on the experiment, serve as a step towards modeling the influence of more general pulse shapes onto the spin dynamics.
We demonstrate that taking into account the finite width has a profound influence on the magnetic field dependent revival amplitude at large external magnetic field.
The electron spin precession period of the order of $10$\,ps in a magnetic field of about $10 \, \rm T$ becomes as short as the laser pulse duration.
In the classical treatment, in turn, a classical approximation of the laser pulses is employed that neglects the trion excitation but, however, respects the quantum uncertainty of the electronic spin components.
The classical approach allows us to treat spin baths of up to $700$ effective nuclear spins calculating pulse sequences up to a million laser pulses in the limit of instantaneous pulses and, hence, corroborates the quantum mechanical results with larger nuclear spin baths.

We present results on the field dependency of the revival amplitude in pump-probe experiments with an expanded range up to $10 \, \mathrm{T}$ for the magnetic field applied perpendicular to the optical axis whereas former experiments had been limited to $6 \, \mathrm{T}$ only~\cite{Jaeschke2017}.
The data for two different (In,Ga)As/GaAs QD samples show a characteristic minimum of the revival amplitude at roughly $4 \, \mathrm{T}$.
Our theoretical approaches disclose that the nuclear Larmor frequency \cite{BeugelingUhrigAnders2017} plays a crucial role to understand these experimental data, e.\,g. $4 \, \mathrm{T}$ roughly corresponds to the external magnetic field where the nuclear spins perform half a revolution between two succeeding pump pulses.
The nuclear Larmor precession determines the electronic resonance condition and, thus, the number of electron spin revolutions between two pump pulses.
Since the number of electron spin revolutions in between two pump pulses also determines the alignment of the electron spin immediately before a pump pulse, we connect the nuclear resonance condition directly to the revival amplitude.

In the experiments, the properties of the QDs, such as the electron $g$-factor and the trion excitation energy, vary in the ensemble.
Detuned QDs are not efficiently pumped and practically do not contribute to the spin polarization.
The mode-locking condition \cite{PhysRevLett.96.227401,GreilichScience2007} in such ensembles, however, causes a synchronization of the electron spin dynamics in periodically pumped QDs with slightly different $g$-factors \cite{PhysRevLett.96.227401,GreilichScience2007,Jaeschke2017}.
In this paper, the theoretical approaches focus on an ensemble with fixed $g$-factor and trion excitation energy, but the quantum mechanical treatment includes variations of the hyperfine coupling accounting for slightly different characteristic dephasing time scales $T^*$ in each QD.

For completeness, we note that there have been extensive studies of the electron-nuclear interaction on the single QD level, see Ref. \cite{Urbaszek2013,Warburton2013,Chekhovich2013}.
The spin coherence of electrons and also holes and in particular its limitation due to coupling to the nuclear bath was studied by echo-type experiments \cite{Press2010,Bechtold2016}.
Requirements for the nuclear spin system to reduce the detrimental effect
on the electron spin coherence were formulated \cite{Chekhovich2015}.
Sophisticated strategies were implemented to suppress the carrier spin dephasing; both in gate-defined QDs \cite{Bluhm2010}, and in self-assembled QDs using pulse sequences for dynamic decoupling \cite{Varwig2014,Huthmacher2018} or coherent population trapping that is sensitive to the nuclear Overhauser field \cite{Issler2010,Prechtel2016}.
Using the latter technique, it was recently possible to monitor the evolution of the nuclear spin bath and to demonstrate an extension of the electron spin dephasing time by an order of magnitude in self-assembled QDs \cite{PhysRevLett.119.130503}.
Vice versa, also the impact of the electron spin on the nuclear spin coherence has been studied \cite{Wuest2016}.
Here we focus on a different problem, namely the contribution of the nuclear spin bath to the synchronization of the electron spin precession about an external magnetic field with the periodically pulsed excitation laser that orients the spins.
We monitor the electron spin coherence in a QD ensemble over times covering eleven orders of magnitude as a function of magnetic field strength.

The paper is organized as follows.
We start by presenting measurements of the revival amplitude obtained in pump-probe experiments in Sec.\ \ref{sec:experiment}.
Then, we turn towards the theoretical calculations.
The CSM underlying both, the quantum mechanical and the classical approach, is introduced in Sec.\ \ref{sec:csm}.
In Sec.\ \ref{sec:qmapproach}, we devise the full quantum mechanical approach to periodically pulsed QDs.
The results obtained by the quantum mechanical approach with instantaneous pump pulses are illustrated in Sec.\ \ref{sec:results-qm-approach}.
These results are compared to the classical approach in Sec.\ \ref{sec:classic-pulses}.
In Sec.\ \ref{sec:pulses}, we extend the quantum mechanical description to pump pulses with Gaussian envelope.
The last section summarizes our theoretical and experimental results and draws conclusions.

\section{Experimental Results}
\label{sec:experiment}

First, we present the experimental results of the magnetic field dependency of the revival amplitude.
We study two different samples of singly charged (In,Ga)As/GaAs QD ensembles using a pump-probe Faraday rotation setup similar to the one in Ref.~\cite{PhysRevLett.96.227401}.
A Ti:Sapphire laser emits pulses of $2.5 \,$ps duration with a repetition period of $13.2 \,$ns.
To polarize the electron spins via trion excitation~\cite{Shabaev2003}, the pump pulses are circularly ($\sigma^{+\text{/}-}$) polarized.
Switching the polarization between $\sigma^{+}$ and $\sigma^{-}$ with a frequency of $84 \,$kHz enables us to perform synchronous detection using a lock-in amplifier.
The samples are cooled to $4.7 \,$K in a cryostat, which is equipped with a superconducting split-coil solenoid and allows us to apply magnetic fields of up to $10 \,$T.
We align the magnetic field perpendicular to the light propagation vector which is parallel to the growth axis of the sample (Voigt geometry).
Directing the probe beam through the sample, the Faraday ellipticity change is detected by an optical differential bridge.

\begin{figure}[t]
  \begin{center}
    \includegraphics[scale=1]{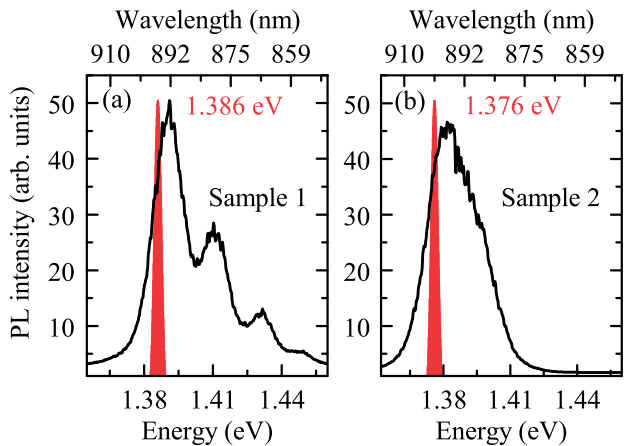}
    \caption{(Color online) Photoluminescence (PL) spectra of the two studied samples. The spectra are taken at a temperature of $4.7 \,$K with a photon excitation energy of $1.631 \,$eV. The laser pulses used in the pump-probe experiment are shown in red. (a) Sample 1 and (b) sample 2.}
    \label{fig:sample_pl}
  \end{center}
\end{figure}

The two samples were grown by molecular-beam epitaxy on a (001)-oriented GaAs substrate.
Each sample features $20$ layers of self-assembled InGaAs QDs with a dot density of approximately $10^{10} \, \text{cm}^{-2}$.
Each QD layer is followed by $16 \,$nm of GaAs.
Then, a Si-donor $\delta$-layer is deposited with a density similar to the QD density, providing therefore on average one electron per dot, so that the QDs are singly charged.
This sheet of donors is followed by a GaAs barrier of $44 \,$nm before the next layer of QDs is grown leading to a total separation of $60 \,$nm between two adjacent dot layers.

After the epitaxial growth, the samples were thermally annealed to homogenize the QD ensembles.
In addition to homogenizing the dot sizes, the annealing also led to a further exchange of Ga and In atoms between the InGaAs QDs and the surrounding GaAs barriers~\cite{Petrov2008} so that the In-content in the QDs is reduced.
Besides, the thermal annealing shifts the emission energy of the sample to higher values.
For both samples the rapid thermal annealing time was chosen to be $30 \,$s. Sample 1 was annealed at $945 \, ^{\circ}$C, sample 2 at $880 \, ^{\circ}$C.
The photoluminescence spectra of both samples as well as the spectra of the exciting pulsed laser in the pump-probe experiments are shown in Fig.~\ref{fig:sample_pl}.
Sample 1 is the sample used in Refs.~\cite{PhysRevLett.96.227401,GreilichScience2006,GreilichScience2007,GREILICH20091466} which we resonantly excite in the low energy flank of the ground state transition at a photon energy of $1.386 \,$eV.
The recombination of electron-hole pairs with the electron in excited QD confined states above the ground state shows up as additional peaks towards higher energies.
Sample 2 has a lower central emission energy and is resonantly excited at a photon energy of $1.376 \,$eV.

\begin{figure}[t]
  \begin{center}
    \includegraphics[scale=1]{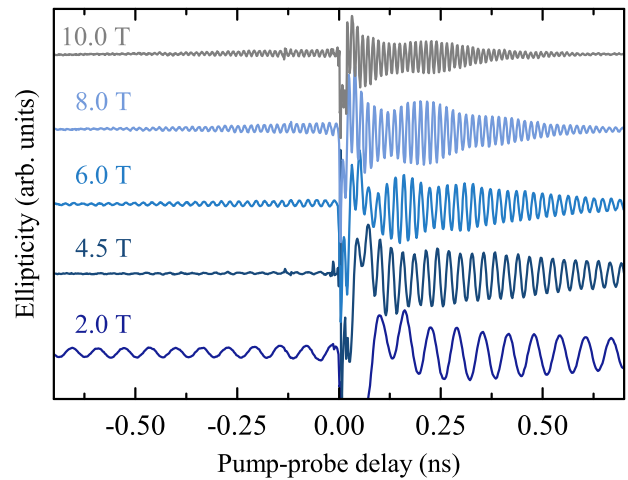}
    \caption{(Color online) Dynamics of the electron spin projection onto the optical axis in different magnetic fields. The Faraday ellipticity measured for sample 2 is plotted versus the pump-probe delay. The dynamics are obtained at a temperature of $4.7 \,$K. The curves are shifted vertically for clarity.}
    \label{fig:exp_dynamics}
  \end{center}
\end{figure}

In the Faraday rotation measurements, the pump and the probe pulse trains were split from the same laser source.
We take pump-probe traces for both samples by incrementing the transit time of the pump pulses through the sample via a mechanical delay line.
These traces are the experimental manifestation of the steady-state spectra of the coupled electron-nuclear system after several million pump pulses in Fig.~\ref{fig:timeevolution}~(a) below.
The signal for each delay step is integrated over $100 \,$ms. Starting from $1 \,$T, we record the dynamics of the electron spin projection onto the optical axis for magnetic fields up to $10 \,$T in steps of $0.5 \,$T.
Fig.~\ref{fig:exp_dynamics} shows a selection of these spectra. At delay $t=0$, the pump pulse aligns the electron spins.
Towards negative and positive delays, the total spin polarization decreases due to a dephasing of the spin ensemble.
The decay of the total spin polarization is superimposed by an oscillating function which reflects the Larmor precession.

We fit the negative side of the spectrum with an inhomogeneously (Gaussian) decaying cosine function (in accordance to Eq.~\eqref{eq:merkulov}):
\begin{align}
S^x(t)=A \cos (\omega t) \exp{\left(-\frac{t^2}{6{T^*}^2} \right)} \, .
\end{align}
From these fits, we can extract the revival amplitude $A$, the Larmor frequency $\omega$ and the dephasing time $T^*$.
The Larmor frequencies show a linear dependence on the magnetic field with an electron $g$-factor of $\ge=0.57$.
The dephasing time $T^*$ obviously decreases with increasing magnetic field. Due to the finite spectral width of the pulses, a distribution of electron $g$-factors is excited that is translated into a spread of precession frequencies. This spread increases with linearly increasing magnetic field, leading to an enhanced dephasing. The measured $T^*$-dependence follows indeed the expected $1/\Bext$-behavior. We note that this dephasing does not impact the discussion of the revival amplitude, as we determine the amplitude right before a pump pulse. 

\begin{figure}[t]
  \begin{center}
    \includegraphics[scale=1]{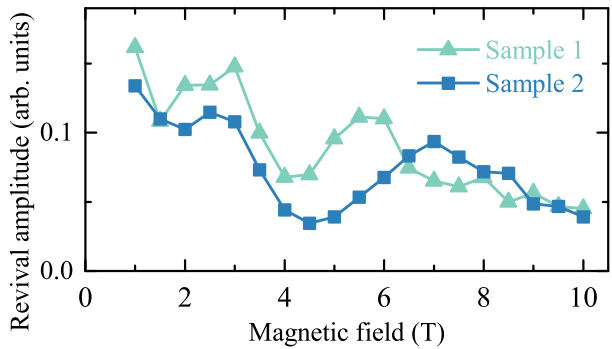}
    \caption{(Color online) Dependence of the revival amplitude on the external magnetic field. The revival amplitudes (the symbols) were extracted from the pump-probe spectra for the two different samples at temperature $4.7 \,$K. The lines are guides to the eye.}
    \label{fig:exp_magn_field}
  \end{center}
\end{figure}

In Fig.~\ref{fig:exp_magn_field}, we plot the revival amplitude in arbitrary units as function of the magnetic field.
Note that the measured revival amplitude depends strongly on the experimental setup and thus cannot be compared quantitatively to the theoretical results in the later sections.

Sample 1 and sample 2 both show a non-trivial magnetic field dependency but with similar characteristics.
The amplitudes decrease towards higher magnetic fields, which we attribute to the varying In and Ga contents in each dot leading to a nuclear $g$-factor spread and a dephasing of the nuclear spins (see Sec.~\ref{sec:revampohf} for further explanation).
Additionally, for both samples we see an oscillatory behavior of the revival amplitude with a main minimum between two maxima, onto which smaller fluctuations are superimposed.
The central minimum for sample 1 is positioned at about $4.2 \,$T, whereas the two maxima occur around $2.5 \,$T and $5.7\,$T, respectively. For sample 2, the maxima and the minimum occur at slightly higher magnetic fields as if the oscillatory period is increased.

\section{Central spin model (CSM)}
\label{sec:csm}

In order to describe the experimental findings, the theoretical approaches in this paper 
target the spin dynamics of a QD ensemble subject to periodic laser pump pulses.
The dynamics are separated into three parts: 
First, the electron spin interacting with its nuclear spin environment is accounted
for by a CSM (also called Gaudin model \cite{Gaudin}).

Hereby, we restrict our description to the hyperfine interaction and neglect other effects such as the nuclear dipole-dipole interaction or the nuclear-electric quadrupolar interaction \cite{CoishBaugh2009,Sinitsyn2012,Bulutay2012,Bechtold2015,hackmannPRL2015}.
The two latter typically are some orders of magnitude smaller than the hyperfine interaction \cite{Dyakonov} and only are relevant on time scales much larger than the pulse repetition time $\TR = 13.2 \, \mathrm{ns}$.
It has been shown \cite{Sinitsyn2012,Bulutay2012,Bechtold2015} 
that the nuclear-electric quadrupolar interactions induce an additional electronic
dephasing time  of the order of $300$\,ns in the absence of an external magnetic field. The effect of this interactions is suppressed
in a finite magnetic field due to its competition with the nuclear Zeeman energy:
 the spin-noise spectrum \cite{Glasenapp2016} can be fitted by a frozen Overhauser field approximation
for external fields exceeding $40$\,mT.
The characteristic dephasing time 
associated with this competing interactions increases with the external field and arrives at values of 2-4\,$\upmu \mathrm{s}$ 
\cite{Press2010,Bechtold2016,FroehlingAnders2017} for $\Bext>3\, \mathrm{T}$. Since the time scale
induced by the nuclear-electric quadrupolar interactions is about $300$ times larger than $\TR$ in a large external magnetic field, 
the nuclear-electric quadrupolar interactions only provide a  small perturbative correction and can be omitted relative to
the leading order effect presented here.

The second ingredient for the spin dynamics in the QD is the light-matter interaction of the classical laser field.
The third part comprises the radiative decay of the laser induced trion state.
At the end, we average over different realizations of QDs to obtain the spin dynamics in a QD ensemble.

The CSM 
\cite{Gaudin,MerkulovEfrosRosen,CoishLoss2004,alhas06,
FischerLoss2008,FaribautSchuricht2013a,FaribautSchuricht2013b}
comprises a bath of $N$ nuclear spins coupled to the electron spin via hyperfine interaction.

The Hamiltonian $\Hcsm$ of the CSM consists of three terms, the hyperfine interaction~$H_{\rm HF}$, the electron Zeeman effect~$H_{\rm EZ}$ and the nuclear Zeeman effect~$H_{\rm NZ}$
\begin{align}
\Hcsm = H_{\rm HF} + H_{\rm EZ} + H_{\rm NZ} \, .
\label{eq:CSM}
\end{align}
These three parts can be written in terms of the electron spin operator $\vec{S}$ and the nuclear spin operators~$\vec{I}_k$
\begin{align}
H_{\rm HF} &= \sum_{k=1}^N \hbar^{-2} A_k \left( S^x I_k^x + S^y I_k^y + S^z I_k^z \right) \notag \\
&= \sum_{k=1}^N \hbar^{-2} A_k \left( S^x I_k^x + \frac12 \left( S^+ I_k^- + S^- I_k^+ \right) \right) \\
H_{\rm EZ} &= \hbar^{-1} \ge \muB \Bext S^x \label{eq:ez}\\
H_{\rm NZ} &= \hbar^{-1} \gN \muN \Bext \sum_{k=1}^N I^x_k \, . \label{eq:nz}
\end{align}

While the negatively charged QDs studied in this paper are described by isotropic coupling constants $A_k$, positively charged QDs require the extension to an anisotropic coupling between electron and nuclear spins \cite{HansonSpinQdotsRMP2007,Testelin2009,HackmannAnders2014}.
Using the spin ladder operators $S^\pm = S^y \pm \im S^z$ and $I_k^\pm = I_k^y \pm \im I_k^z$ the hyperfine interaction can be rewritten as an Ising term parallel to the external magnetic field (in $x$-direction) and two spin-flip terms.

In the Zeeman terms, $\ge$ and $\gN$ denote the electron and the nuclear $g$-factor, respectively.
The constants $\muB$ and $\muN$ are the Bohr magneton and the nuclear magneton.
Note that we choose one effective value $\gN \muN$ for all nuclei.
Different types of nuclei have been treated for instance in Ref.~\cite{BeugelingUhrigAnders2017}, but are beyond the scope of the present work.
In our calculations, we use an electron $g$-factor $\ge = 0.555$, which is typical in experimental studies of InGaAs QDs \cite{GREILICH20091466, GreilichScience2007}.
This leads to an angular electron Larmor frequency $\omega_e = \muB \ge \Bext / \hbar$ of roughly 
$97.6 \cdot 10^9 \, \mathrm{rad/s}$ at $\Bext = 2 \, \mathrm{T}$.
For the nuclear spins, we choose a $800$ times slower precession with the ratio $z = \gN \muN / (\ge \muB ) = 1 / 800$.
This value is based on the weighted average of the nuclear magnetic moments of Ga and As \cite{Walchli,Stone} and has been calculated in Ref.~\cite{BeugelingUhrigAnders2017}.
Thus, the nuclear angular Larmor frequency $\omega_N = \muN \gN \Bext / \hbar$ is roughly $122 \cdot 10^6 \, \mathrm{rad/s}$
at $\Bext = 2 \, \mathrm{T}$ and $\gN \approx 1.27$.

The Hamiltonian $\Hcsm$ is diagonal in the spin $x$-basis except for the spin-flip terms in $H_{\rm HF}$.
In the following, we denote the electron spin $x$-basis, {i.\,e.}, the eigenbasis of $H_{\rm EZ}$, by $\ket{\up}$ and $\ket{\down}$.
Therefore, it is $S^x \ket{\up} = \hbar / 2 \ket{\up}$ and $S^x \ket{\down} = - \hbar / 2 \ket{\down}$.
For the sake of simplicity, we also treat the nuclear spins as spins $1/2$ even though in real QDs the nuclei have spin $3/2$ (Ga and As) and spin $9/2$ (In) \cite{Walchli,Stone}.
The assumption of nuclear spins $1/2$ restricts the dimension of the density matrix in our quantum mechanical approach to $2\cdot 2^N$ with two spin states for the electron and each nucleus, respectively.

The hyperfine coupling constants $A_k$ arise from the Fermi contact interaction.
Therefore, their values are determined by the electron wave function $| \psi ( \vec{R}_k ) |^2$ at the position of a nucleus \cite{MerkulovEfrosRosen}.
The hyperfine interaction $H_{\rm HF}$ can be interpreted in terms of an additional magnetic field that acts on the electron spin and is caused by the nuclear spins.
This additional magnetic field is called the Overhauser field
\begin{align}
\vec{B}_{N} = \left( \ge \muB \hbar \right)^{-1} \sum_{k=1}^N A_k  \vec{I}_k \, .
\label{eq:ohf}
\end{align}
The additional magnetic field that is caused by the electron spin and acts on nuclear spin $k$, in turn, is termed Knight field
\begin{align}
\label{eq:knight-field}
\vec{B}_{k,\mathrm{Kn}} = \left( \gN \muN \hbar \right)^{-1} A_k \vec{S}.
\end{align}
The fluctuation of the Overhauser field $\vec{B}_N$ leads to a dephasing of the electron spin with a characteristic time~$T^*$~\cite{MerkulovEfrosRosen}
\begin{align}
\left( T^* \right)^{-2} = \hbar^{-4} \sum_{k=1}^N A_k^2 \expect{I_k^2} \, .
\label{eq:tstar_noensemble}
\end{align}
In the experiment, the dephasing time typically is of the order of $1$ to $3$ nanoseconds~\cite{GreilichScience2006,GreilichScience2007,PhysRevLett.96.227401} 
if fitted proportional to $\exp \left( - t^2/(T^*)^2 \right)$ as in Ref.\ \cite{MerkulovEfrosRosen}.
The definition of $T^*$ in Eq.~\eqref{eq:tstar_noensemble}, however, leads to a dephasing with envelope~\eqref{eq:merkulov} such that $T^*$ takes values in the range of $0.4$ to $1.2$ nanoseconds.
These experimental values include additional dephasing mechanisms, {e.\,g.}, the electron $g$-factor spread as discussed above. 
Since the two theoretical approaches in this paper only comprise the electron dephasing due to the hyperfine interaction, we adjust the characteristic dephasing time $T^*$ to the experimental values of the overall dephasing time mimicking other effects as well.

\section{Quantum mechanical approach to periodically pulsed QDs}
\label{sec:qmapproach}

The scope of this work is to calculate the approach of the spin dynamics to steady-state in a periodically driven QD ensemble.
In order to access this limit numerically with a full quantum mechanical simulation, several million pump pulses are required.
Since the computational time grows exponentially with the Hilbert space dimension, we restrict our calculation to a rather small bath size of $N=6$ nuclear spins. 

\subsection{Hyperfine coupling constants $A_k$}
\label{sec:ak}

\begin{figure}[htbp]
\begin{center}
\includegraphics[scale=1]{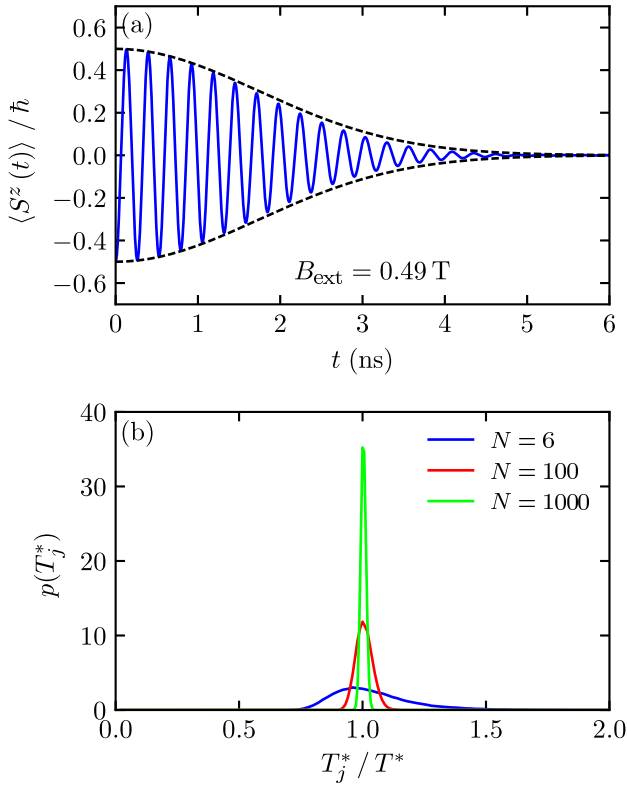}
\caption{(Color online) Influence of the hyperfine coupling constants $A_k$. (a) The numerically calculated time evolution of the electron spin component $\Szt$. At $t=0$ the spin is directed in negative $z$-direction. The Gaussian envelope function according to Eq.~\eqref{eq:merkulov} is indicated by black dashed lines. Small deviations are due the finite number of nuclei $N=6$ ($\NC = 100$). (b) The distribution $p(T^*_j)$ of the dephasing time $T^*_j$ within a single configuration for different numbers $N$ of nuclei. Here, we scale the coupling constants to $T^* = 1 \, \mathrm{ns}$ in $1000$ sets with $\NC = 100$ configurations each, in order to obtain an approximately continuous distribution $p(T^*_j)$.}
\label{fig:tstar}
\end{center}
\end{figure}

A real QD typically contains of the order $10^5$ nuclear spins with couplings $A_k$ that are given by a distribution function $p(A_k)$.
It has already been shown that a representation of the system with a reduced number of nuclear spins is able to reproduce the generic spin dynamics of a larger system \cite{HackmannAnders2014}.
To compensate for the small number of nuclear spins and to simultaneously include fluctuations induced by the slightly different QDs in the ensemble, we consider $\NC=100$ realizations of the CSM.
These realizations differ in their set of hyperfine coupling constants $\{A_{k}\}$.
During the whole pulse sequence the configurations are treated as independent representations of a single QD and the results are merged only at the end.
As a side product, the computation time scales only linearly with $\NC$.
To distinguish the configurations, we introduce an index $j \in \{ 1, ... , \NC \}$, {e.\,g.}, $A_{k,j}$ is the coupling constant for nuclear spin $k$ in configuration $j$.
For brevity, the index $j$ will be omitted, when we consider a single configuration only.

Since the details of the distribution function have a weak influence on the steady-state dynamics \cite{Jaeschke2017}, we choose the coupling constants uniformly distributed in the range $\left[ 0.2 ; 1 \right]$.
In this way, we exclude very small couplings to nuclear spins, which have minor impact on the electron spin.
The randomly distributed coupling constants $A_{k,j}$ lead to an ensemble averaged dephasing time
\begin{align}
\left( T^* \right)^{-2} = \frac{1}{\NC \, \hbar^4} \sum_{j=1}^{\NC} \sum_{k=1}^N A_{k,j}^2 \expect{I_{k,j}^2} \, ,
\label{eq:tstar}
\end{align}
where $\big<I_{k,j}^2\big> = \expect{I^2} = \frac34 \hbar^2$ for nuclear spins $1/2$.
In our calculations, we set $T^* = 1 \, \rm ns$ based on the experiments and scale the coupling constants accordingly.

Fig.~\ref{fig:tstar}~(a) shows the time evolution of the electron spin component $\Szt$ calculated by averaging all $\NC$ configurations (see Eq.~\eqref{eq:si}).
Small deviations from the Gaussian envelope function \cite{MerkulovEfrosRosen}
\begin{align}
\Szt_\mathrm{env} = S_0 \, \exp \left( - \frac{t^2}{6 \, {T^*}^2} \right)
\label{eq:merkulov}
\end{align}
are caused by the finite number of $N=6$ nuclear spins.

Since we fix the ensemble averaged dephasing time in Eq.~\eqref{eq:tstar}, $T^*$ varies in the different configurations $j$ mimicking an ensemble of quantum dots.
This variation is depicted in Fig.~\ref{fig:tstar}~(b) for different numbers of nuclei $N$.
For that purpose, we define the ratio $T^*_j/T^*$ via
\begin{align}
\left(\frac{T^*}{T^*_j} \right)^{2} = \frac{1}{\hbar^4}  \sum_{k=1}^N (T^*A_{k,j})^2 \expect{I_{k,j}^2}
\end{align}
for each of the $\NC=100$ configurations entering the definition of $T^*$ in Eq.\ \eqref{eq:tstar}.
The distribution $p (T^*_j)$ is obtained from $1000$ such sets containing $\NC=100$ configurations each.
The distribution $p (T^*_j)$ clearly reveals a self-averaging effect for increasing $N$ if normalized $a_{k,j} =  T^*A_{k,j}$ are used.
For simulations with a large number of nuclei $N$, one has to replace our approach by $a_{k,j}  =  T^*_jA_{k,j}$, where $T^*_j$ must be randomly generated from a distribution $p (T^*_j)$ with a fixed width corresponding to the experimental variations of the QDs.

\subsection{Instantaneous laser pump pulses}
\label{sec:instpp}

In order to describe the time evolution during a laser pump sequence, the CSM has to be extended by the trion state $\ket{\rm T} = \ket{ \up \down \Uparrow}_x$, which is excited by the circularly polarized pump pulses \cite{PhysRevB.85.125304}.
Since we consider $\sigma^+$-polarized light only, we omit the trion state $\ket{ \up \down \Downarrow}_z$.
Hence, the electronic subspace comprises three possible states, {i.\,e.}, $\{ \up_z, \down_z, \mathrm{T} \}$, and the full density matrix has dimension $\left( 3 \cdot 2^N \right) \times \left( 3 \cdot 2^N \right)$.
Here, we choose the spin basis along the optical axis in $z$-direction.
The states $\ket{\up}_z$ and $\ket{\down}_z$ can be transformed into the magnetic field eigenbasis, $\ket{\up}$ and $\ket{\down}$, via ${\ket{\up}_z = \left( \ket{\up} + \ket{\down} \right) / \sqrt{2}}$ and $\ket{\down}_z = \left( \ket{\up} - \ket{\down} \right) / \sqrt{2}$, respectively.

At first, we treat the laser pulses in the limit of vanishing duration, hence considering them as instantaneous.
Later, in Sec.\ \ref{sec:pulses}, we will lift this approximation.
The impact of an instantaneous $\pi$-pulse, which resonantly excites the trion state, is given by $\rho \rightarrow U_P \rho U_P^\dag$ with the unitary pulse operator
\begin{align}
U_P = \ket{\rm T} \bra{\up}_z - \ket{\up}_z \bra{\rm T} + \ket{\down}_z \bra{\down}_z \, .
\end{align}
This unitary transformation of the density operator $\rho$ corresponds to a complete exchange of the $\ket{\up}_z$ population and the $\ket{\rm T}$ population.
Meanwhile, the $\ket{\down}_z$ population remains unaffected by the pulse.
Note that the pulse operator $U_P$ does not have any effect on the nuclear spin configurations at all.

\subsection{Lindblad approach}
\label{sec:lindblad}

Due to the trion decay after each pump pulse, a unitary time evolution between pulses would have to include the participating photons.
Since we are not interested in the resonance fluorescence, we treat the trion decay in the framework of an open quantum system, {i.\,e.}, by a master equation in Lindblad form \cite{Carmichael} for the time evolution of the density operator $\rho$ between two succeeding pump pulses
\begin{align}
\frac{\mathrm{d} \rho}{\mathrm{d} t} = - \frac{\im}{\hbar} \left[ H , \rho \right] + \gamma \left( s^\dag s \rho + \rho s^\dag s - 2 s \rho s^\dag \right) = \mathcal{L} \rho
\label{eq:lb}
\end{align}
and treat the photon emission by a spontaneous Markov process with rate $\gamma$.
The term including the commutator of the von Neumann equation contains the unitary part of the time evolution, namely the spin dynamics captured by the CSM.
Here, Hamiltonian $H$ includes $\Hcsm$ and the trion state with excitation energy $\e$
\begin{align}
H = \Hcsm + \e \ket{{\rm T }} \bra{{\rm T}} \, .
\end{align}
The term proportional to $\gamma$ in the Lindblad equation accounts for the trion decay.
The decay rate $\gamma$ is set to $10 \, \mathrm{ns}^{-1}$ based on experimental data for a trion lifetime of about $400 \, \mathrm{ps}$ \cite{PhysRevLett.96.227401}.
The operators $s = \ket{\up}_z \bra{\mathrm{T}}$ and $s^\dag = \ket{\mathrm{T}} \bra{\up}_z$ map the trion state to the spin-up state along the optical axis and vice versa.
The whole time evolution of $\rho$ can be written in terms of a super operator, the so called Liouville operator $\mathcal{L}$.
For a time independent $\mathcal{L}$, the solution to Eq.~\eqref{eq:lb} is given by an exponential function
\begin{align}
\rho \left( t \right) = {\rm e}^{- \mathcal{L} t} \rho \left( 0 \right),
\end{align}
which is valid for the times between two pulses, where $\rho \left( 0 \right)$ is the density operator right after the pulse.
However, the actual calculation of this solution would involve diagonalization of the matrix representation of $\mathcal{L}$, which has dimension $\left( 3 \cdot 2^N \right)^2 \times \left( 3 \cdot 2^N \right)^2$.
In order to circumvent this time-consuming task, we develop an alternate approach that is described below.
Within this method, we only have to treat matrices of much smaller dimension $\left( 2 \cdot 2^N \right) \times \left( 2 \cdot 2^N \right)$.

\subsection{Partitioning of the density operator}
\label{sec:implementation}

To solve the Lindblad equation, we first transform into the frame rotating with the Larmor frequency $\omega_{N}$ of the nuclear spins.
Hereby, we eliminate the nuclear Zeeman term in the Hamiltonian.
The transformed Lindblad equation reads
\begin{align}
\dot{\tilde{\rho}} = \- \frac{\im}{\hbar} \left[ \tilde{H} , \tilde{\rho} \right] - \gamma \left( \tilde{s}^\dag \tilde{s} \tilde{\rho} + \tilde{\rho} \tilde{s}^\dag \tilde{s} - 2 \tilde{s} \tilde{\rho} \tilde{s}^\dag \right) \, ,
\end{align}
where the transformed operators $\tilde{O} = U_{\rm RF} O U_{\rm RF}^\dag$ are denoted by "$\, \tilde{\;} \,$" and 
\begin{align}
U_{\rm RF} = \exp \left\{ - \im \; \frac{\omega_{N}}{\hbar} \left( S^x + \sum_k I_k^x \right) t \right\} \, .
\end{align}
The new Hamiltonian in the rotating frame is given by
\begin{align}
\tilde{H} = \left( \omega_{e} - \omega_{N} \right) S^x + H_{\rm HF} + \e \ket{\mathrm{T}} \bra{\mathrm{T}} \, .
\label{eq:hrf}
\end{align}

In this frame, the electron precesses with the reduced frequency $\omega_{e} - \omega_{N}$, while the hyperfine interaction remains unaffected by the transformation.
The operator $\tilde{s}$ in the rotating frame, defined in the basis along the external magnetic field, is
\begin{align}
\tilde{s}=\frac{1}{\sqrt{2}} \left( \mathrm{e}^{- \im \omega_{N} t / 2} \ket{\up} + \mathrm{e}^{\im \omega_{N} t / 2 } \ket{\down} \right) \bra{\rm T} \, .
\end{align}
Inserting $\tilde{s}$ and its conjugate $\tilde{s}^\dag$, the Lindblad equation yields
\begin{align}
\dot{\tilde{\rho}} = &- \frac{\im}{\hbar} \left[ \tilde{H} , \tilde{\rho} \right] - \gamma \Big( \ket{\mathrm{T}} \bra{\mathrm{T}} \tilde{\rho} + \tilde{\rho} \ket{\mathrm{T}} \bra{\mathrm{T}} \Big) \notag \\
&+ \gamma \braket{\mathrm{T}}{\tilde{\rho}}{\mathrm{T}} \Big( \ket{\up}\bra{\up} + \ket{\down} \bra{\down} \Big. \notag \\
&+ \Big. \mathrm{e}^{-\im \omega_{N} t} \ket{\up}\bra{\down} + \mathrm{e}^{\im \omega_ {N} t} \ket{\down}\bra{\up} \Big) \, .
\label{eq:lbrf}
\end{align}

This Lindblad equation in the rotating frame allows us to separate the trion decay from the remaining dynamics.
In the electron-nuclear tensor space spanned by the basis $\ket{e,K}$, one can define a reduced density operator $\tilde{\rho}_{\rm TT} = \braket{\mathrm{T}}{\tilde{\rho}}{\mathrm{T}}$ acting only on the nuclear spin configurations $\ket{K}$, while the electronic state $e \in \{ \up, \down \rm T \}$ has been fixed to the trion state $\rm T$.
Apparently, the dynamics of this operator obeys
\begin{align}
\dot{\tilde{\rho}}_{\rm TT} = - 2 \gamma \tilde{\rho}_{\rm TT}
\label{eq:rhott}
\end{align}
and its matrix representation has the dimension $2^N \times 2^N$ determined from the nuclear Hilbert space only.
The analytic solution to Eq.~\eqref{eq:rhott} is an exponential decay of the trion population for arbitrary nuclear spin configurations
\begin{align}
\label{eq:18}
\tilde{\rho}_{\rm TT} \left( t \right) = \tilde{\rho}_{\rm TT} \left( 0 \right) \mathrm{e}^{-2 \gamma t}
\end{align}
that decouples from the electronic subsystem.
Therefore, there is no nuclear dynamics in this sector of the density matrix.

We partition the density operator into the remaining eight reduced density operators
$\tilde{\rho}_{ e,e'} = \braket{e}{\tilde{\rho}}{e'}$ and first focus on the four contributions involving the trion, namely the trion coherence sub operators $\braket{\mathrm{T}}{\tilde{\rho}}{\up}$, $\braket{\mathrm{T}}{\tilde{\rho}}{\down}$, $\braket{\up}{\tilde{\rho}}{\mathrm{T}}$ and $\braket{\down}{\tilde{\rho}}{\mathrm{T}}$.
Their differential equations, which we obtain from Eq.~\eqref{eq:lbrf}, decouple from those of the sub operators without trion.
As a result, the elements of trion coherence sub operators decay exponentially with $\gamma$ and we do not have to further investigate them since $\gamma \TR \gg 1$.

We now concentrate on the time evolution for the sub operator $\tilde{\rho}_{\rm S}$ comprising the four reduced density operators involving no trion.
The matrix representation of $\tilde{\rho}_{\rm S}$ has the dimension $\left( 2 \cdot 2^N \right) \times \left( 2 \cdot 2^N \right)$ and only contains the spin-up and spin-down state for the electron.
We insert the analytic solution for the operator $\tilde{\rho}_{\rm TT}(t)$, {i.\,e.}, Eq.~\eqref{eq:18}, into the Lindblad equation~\eqref{eq:lbrf} to determine the time evolution of $\tilde{\rho}_{\rm S}$:
\begin{align}
\dot{\tilde{\rho}}_{\rm S} + \frac{\im}{\hbar} \left[ \tilde{H}_{\rm S} , \tilde{\rho}_{\rm S} \right] = 
& \gamma \tilde{\rho}_{\rm TT} \left( 0 \right) \mathrm{e}^{-2 \gamma t} \Big( \ket{\up}\bra{\up} + 
\ket{\down} \bra{\down} \Big. \notag \\
& \hspace{-1cm} + \Big. \mathrm{e}^{-\im \omega_{N} t} \ket{\up}\bra{\down} + 
\mathrm{e}^{\im \omega_{N} t} \ket{\down}\bra{\up} \Big) \, ,
\label{eq:lbrhos}
\end{align}
where $\tilde{H}_{\rm S}=  \left( \omega_{e} - \omega_{N} \right) S^x + H_{\rm HF} $ is the projection of $\tilde{H}$ onto the spin-spin subspace.
The differential equation for $\tilde{\rho}_{\rm S}$ was divided into the homogeneous part on the left hand side and a source term stemming from the trion decay on the right hand side of the equation.
It can be solved by combining the solution for the homogeneous part of the equation and a particular solution for the full inhomogeneous equation.
Since the homogeneous part equals a von Neumann equation, it is solved by unitary time evolution
\begin{align}
\tilde{\rho}_{\rm S,h} \left( t \right) = \mathrm{e}^{- \im \tilde{H}_{\rm S} t / \hbar} \tilde{\rho}_{\rm S,h} \left( 0 \right) \mathrm{e}^{\im \tilde{H}_{\rm S} t/ \hbar} \, .
\label{eq:rhoh}
\end{align}
A particular solution to the full equation can be obtained by the ansatz
\begin{align}
\tilde{\rho}_{\rm S,nh} \left( t \right) = & \tilde{\chi}_0 \mathrm{e}^{-2 \gamma t} + \tilde{\chi}_+ \mathrm{e}^{\left( \im \omega_{\rm N} - 2\gamma \right) t} \notag \\
&+ \tilde{\chi}_- \mathrm{e}^{\left( - \im \omega_{\rm N} - 2 \gamma \right) t} \, .
\label{eq:rhonh}
\end{align}
For further details on the numerical calculation of $\tilde{\chi}_0$, $\tilde{\chi}_+$  and $\tilde{\chi}_-$, see Appendix~\ref{app:partsol}.
Finally, the solution to the Lindblad Eq.~\eqref{eq:lbrhos} is given by
\begin{align}
\tilde{\rho}_\mathrm{S} \left( t \right) = \tilde{\rho}_\mathrm{S,h} \left( t \right) + \tilde{\rho}_\mathrm{S,nh} \left( t \right) \, ,
\label{eq:timeevolrhos}
\end{align}
where the initial condition directly after a pump pulse at $t=0$ yields $\tilde{\rho}_\mathrm{S,h} \left( 0 \right) = \tilde{\rho}_\mathrm{S} \left( 0 \right) - \left( \tilde{\chi}_0 + \tilde{\chi}_+ + \tilde{\chi}_- \right)$.

To evaluate the solution just before the next laser pulse at $t=\TR$, the contribution of $\tilde{\rho}_\mathrm{S,nh} \left( \TR \right)$ can be neglected due to the exponential decay with decay rate $\gamma$, since $\gamma \TR \gg 1$.
Therefore, the effect of the trion decay is a correction of the density operator $\tilde{\rho}_\mathrm{S} \left( 0 \right)$ in the electronic sector right after the pulse into $\tilde{\rho}_\mathrm{S,h} \left( 0 \right)$, which allows the calculation of the time evolution until the next pulse by a single unitary transformation substituting $t\to \TR$ in Eq.~\eqref{eq:rhoh}.

Iterating the elementary building block, that combines the effect of a single instantaneous pump pulse and the time evolution for $\TR$, we calculate the effect of pulse sequences with up to $20$ million laser pulses.
Note that we present an exact approach to the spin dynamics of a QD subject to sequential pulses for a finite nuclear spin bath.
In contrast, the perturbative approach presented in Ref.~\cite{BeugelingUhrigAnders2016} only
includes up to one spin flip of the nuclear spin system between subsequent pulses.
Our approach is limited to a smaller nuclear spin bath sizes but allows for the simulation of $\NP > 10^7$ pump pulses, while the approach in Ref.~\cite{BeugelingUhrigAnders2016} was restricted to approximately $10^4$ pulses due to CPU run time limitations.

\section{Results of the quantum mechanical approach}
\label{sec:results-qm-approach}

In this section, we present results for the time evolution of the electron spin polarization along the optical axis and in particular the electron spin revival amplitude obtained by the quantum mechanical approach.
We explicitly make contact to the non-monotonic magnetic field dependency of electron spin revival amplitude
found in the experiment and presented in Fig. \ref{fig:exp_magn_field}.
Additionally, we can directly access the nuclear spins in the numerical calculations, which, in contrast, is impossible in the pump-probe experiments.
As a signature of the nuclear state, we present a detailed analyzes of the distribution of Overhauser fields.

\subsection{Evaluation of numerical results}
\label{sec:ohf}

We calculated the quantum mechanical time evolution for each single configuration of a QD represented by a fixed but random selection of $\{A_k\}$.
After iterating pump pulse and time evolution of duration $\TR$ up to the desired number of laser pulses, the average over many calculations of this sort describes the ensemble of QDs.
In this way, the electronic expectation value of the spin polarization is given by
\begin{align}
\expect{S^i \left( t \right)} = \NC^{-1} \sum_{j=1}^{\NC} \expect{S^i \left( t \right)} _j \, ,
\label{eq:si}
\end{align}
where $\expect{S^i \left( t \right)} _j = \Tr{ S^i \rho_j \left( t \right)}$ with $i \in \{ x,y,z \}$ denotes the quantum mechanical expectation value in configuration $j$ at time $t$.

In addition to the electron spin, we are also interested in the effect of the pump sequence on the alignment of the nuclear spins.
The distribution of $\bnx$ along the external magnetic field axis \cite{PetrovYakovlev2012,BeugelingUhrigAnders2016,Jaeschke2017} as defined in Eq.~\eqref{eq:ohf} can be obtained from the configuration average
\begin{align}
\bnx &= \NC^{-1} \sum_{j=1}^{\NC} \expect{\bnxj} = \NC^{-1} \sum_{j=1}^{\NC} \Tr{\bnxj \rho_j} \notag \\
&= \NC^{-1} \sum_{j=1}^{\NC} \sum_{e,K} \braket{e, K}{\rho_j}{e, K} B_{K,j}^x \, ,
\label{eq:ofd}
\end{align}
where $e \in \{ \up , \down , \mathrm{T} \}$ labels the electron degree of freedom and $K$ denotes the configuration of the $N$ nuclear spins with quantization axis in $x$-direction.

Eq.~\eqref{eq:ofd} can be interpreted such that the value
\begin{align}
B_{K,j}^x = \braket{K}{\bnxj}{K}
\end{align}
occurs with probability
\begin{align}
p_{K,j} = \NC^{-1} \braket{e, K}{\rho_j}{e, K} \, .
\label{eq:pKj}
\end{align}
Accumulating all probabilities for a fixed value $\bnx$,
\begin{align}
p \left( \bnx \right) = \frac{1}{\NC} \sum_{j=1}^{\NC} \sum_{e,K} \braket{e, K}{\rho_j}{e, K} 
\delta( \bnx- B_{K,j}^x)
\end{align}
defines the continuous distribution $p \left( \bnx \right)$ for $\bnx$ \cite{BeugelingUhrigAnders2017}, whose integral is normalized to unity by construction \cite{PetrovYakovlev2012}.

\subsection{Electron spin revival amplitude}
\label{sec:elrevamp}

\begin{figure}[b]
\begin{center}
\includegraphics[scale=1]{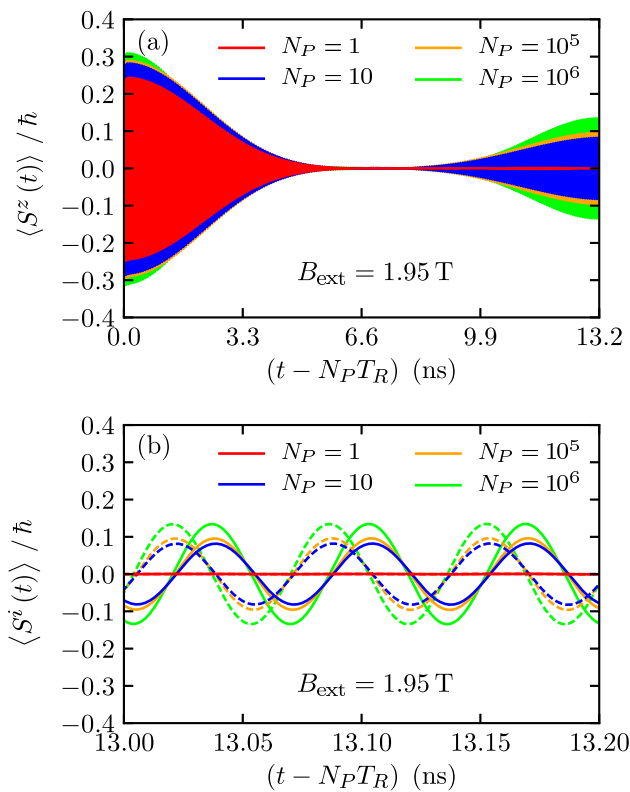}
\caption{(Color online) (a) Time evolution of the electron spin component $\Sz$. Various colors show the time evolution after different numbers $\NP$ of pump pulses. The time axis starts with the arrival of the $\NP$-th pump pulse and ends before the arrival of the next pump pulse after the repetition time $\TR = 13.2 \, \mathrm{ns}$. (b) Time evolution directly before the next pump pulse. In addition to the electron spin component $\Sz$ (solid lines), the electron spin component $\expect{S^y}$ is depicted (dashed lines) to show the spin precession.}
\label{fig:timeevolution}
\end{center}
\end{figure}

\begin{figure}[b]
\begin{center}
\includegraphics[scale=1]{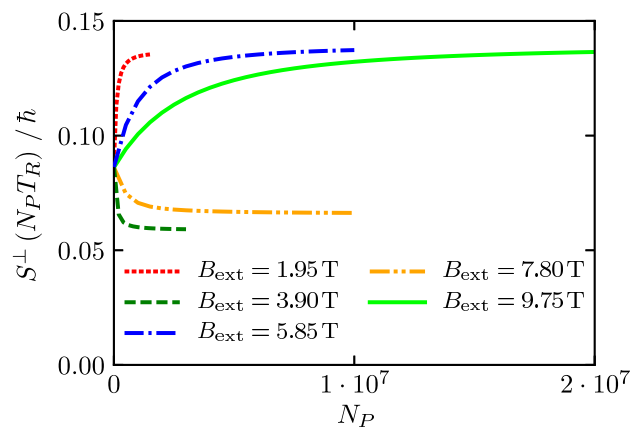}
\caption{(Color online) Evolution of the electron spin revival amplitude $S^\perp \left( \NP \TR \right)$ with the pulse number $\NP$. Various colors show the development for different external magnetic fields $\Bext$.}
\label{fig:revampnp}
\end{center}
\end{figure}

First, we investigate the effect of a sequence of pump pulses with separation $\TR = 13.2 \, \mathrm{ns}$ on the electron spin polarization along the optical axis ($z$-direction).
In Fig.~\ref{fig:timeevolution}~(a), the evolution of $\Sz$ in the time interval between two pulses is depicted for different numbers $\NP$ of applied pump pulses.
The dephasing after the first pump pulse (red curve) is approximately a Gaussian and determined by the dephasing time $T^*$ defined in Eq.~\eqref{eq:tstar}.
The time evolution of $\Sz$ after a large number of pump pulses corresponds to the experimental measurements in Fig.~\ref{fig:exp_dynamics}.
Note that the fast Larmor oscillation in the external magnetic field $\Bext = 1.95 \, \mathrm{T}$ is not resolved on the time scale in Fig.~\ref{fig:timeevolution}~(a) leading to the colored areas.
In Fig.~\ref{fig:timeevolution}~(b), the electron Larmor oscillation of both transversal spin components directly before the arrival of the next pump pulse is presented:
The $y$-component almost vanishes for pumping with instantaneous ideal $\pi$-pulses.

After roughly ten pump pulses, a revival of spin polarization has established with a maximum just before the next pump pulse.
The amplitude of this initial revival (approximately $0.077$) is independent of the external magnetic field, since it originates from a purely electronic steady-state (see Appendix~\ref{app:elss}) \cite{BeugelingUhrigAnders2017}.
We define the revival amplitude as the spin polarization
\begin{align}
\SpTR = \sqrt{ \SzTR ^2 + \left< S^y \left( \NP \TR \right) \right> ^2}
\label{eq:sperp}
\end{align}
after $\NP$ pulses right before the next pulse.
This definition is motivated by the experimental procedure, where an envelope function is fitted to the measured oscillating signal in order to obtain the revival amplitude (see Sec.~\ref{sec:experiment}).
For the numerical calculations, this procedure is not necessary, as we can directly read off the amplitude via Eq.~\eqref{eq:sperp}.

Starting from the initial value originating from the purely electronic steady-state, the revival amplitude evolves further upon increasing the number of pump pulses.
This evolution, however, is dependent on the external magnetic field.
The growth of revival amplitude for $\Bext = 1.95 \, \mathrm{T}$ is shown in Fig.~\ref{fig:revampnp} (red curve).
In addition, the evolution of the revival amplitude for other external magnetic fields is pictured.
For distinct magnetic fields, an increase or a decrease of amplitude with the number of pump pulses $\NP$, can be observed.
But, the rate of change with $\NP$ becomes much slower compared to the initial revival obtained after ten pulses, especially for stronger external magnetic fields.
The magnetic field dependency of the revival amplitude results from a synchronization of the dynamics of all spins including the nuclei with the pump pulses, {{i.\,e.}, from the nuclear focusing.
The periodically pulsed electron spin transmits the effect of the pump pulses to the nuclear spins via the hyperfine coupling.
Therefore, the nuclear spins gradually align along the external magnetic field, which in turn focuses the electron Larmor frequency and thereby leads to either an amplification or a reduction of the initial revival.

\begin{figure}[b]
\begin{center}
\includegraphics[scale=1]{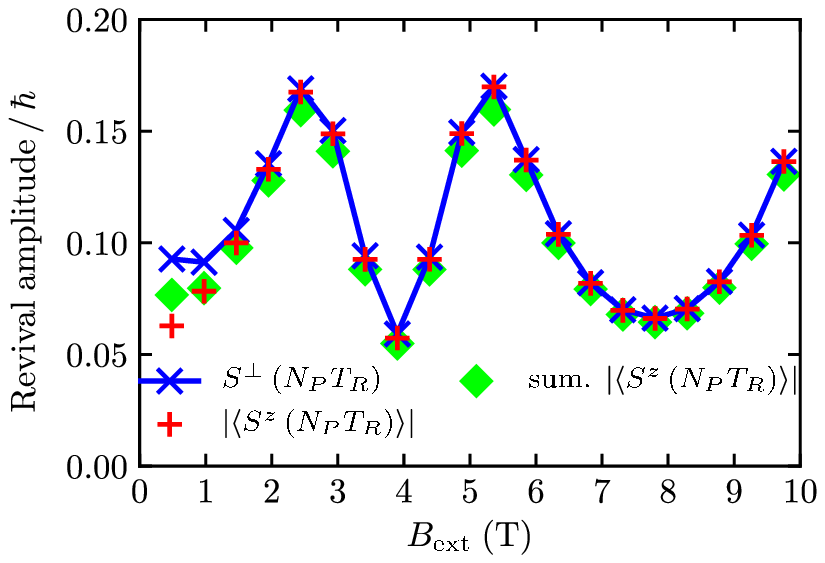}
\caption{(Color online) Magnetic field dependency of the electron spin revival amplitude. 
The converged revival amplitude $\SpTR$ after $1.5 \cdot 10^6 \leq \NP \leq 20 \cdot 10^6$ is depicted as the blue curve.
The exact number of pump pulses depends on the magnetic field $\Bext$.
The dominating electron spin component $\SzTR$ in the full expression for $\SpTR$ is indicated by red crosses.
Furthermore, the revival amplitudes calculated from the Overhauser field distributions according to Eq.~\eqref{eq:sxtr} are added as green diamonds.}
\label{fig:revampb}
\end{center}
\end{figure}

To analyze the magnetic field dependency of the revival amplitude $\SpTR$ in more detail, we plot the converged revival amplitude after up to $20$ million pump pulses as function of the external magnetic field.
The result in Fig.~\ref{fig:revampb} (blue curve) shows a non-monotonic behavior with maxima at approximately $2 \, \mathrm{T}$ and $6 \, \mathrm{T}$ and minima at $4 \, \mathrm{T}$ and $8 \, \mathrm{T}$, respectively.
This behavior can also be observed in the spin component $\SzTR$ (red crosses), which matches $S^\perp \left( \NP \TR \right)$ for external magnetic fields above $2\,$T.
Therefore, we observe that the contribution of the spin component $\SyTR$ nearly vanishes at the incidence time of a pump pulse, which was already indicated in Fig.~\ref{fig:timeevolution}~(b).
The non-monotonic behavior of $\SpTR$ and  $\SzTR$ is caused by the resonance of the nuclear spins, which depends on the external magnetic field and is investigated in the next section by means of the Overhauser field distribution.

Compared to the experimental results in Fig.~\ref{fig:exp_magn_field}, the revival amplitude shows a more pronounced oscillatory behavior demonstrating two equally pronounced maxima in the magnetic field range up to $10 \, \rm T$.
The results have a minimum in common at roughly $4\, \rm T$ enclosed by the maxima at lower and higher external magnetic fields.
In the experiments, the amplitude of the maxima decreases with stronger external magnetic fields.
This effect is not visible in Fig.~\ref{fig:revampb} indicating that certain aspects are not yet captured.
Among these are (i) some sample dependencies as depicted in Fig.~\ref{fig:exp_magn_field} (see Sec.~\ref{sec:revampohf}) as well as (ii) the approximation of an instantaneous pump pulse  which is inappropriate for larger magnetic fields (see Sec.\ \ref{sec:pulses}).

\subsection{Overhauser field distribution}
\label{sec:ofd}

\begin{figure}[b]
\begin{center}
\includegraphics[scale=1]{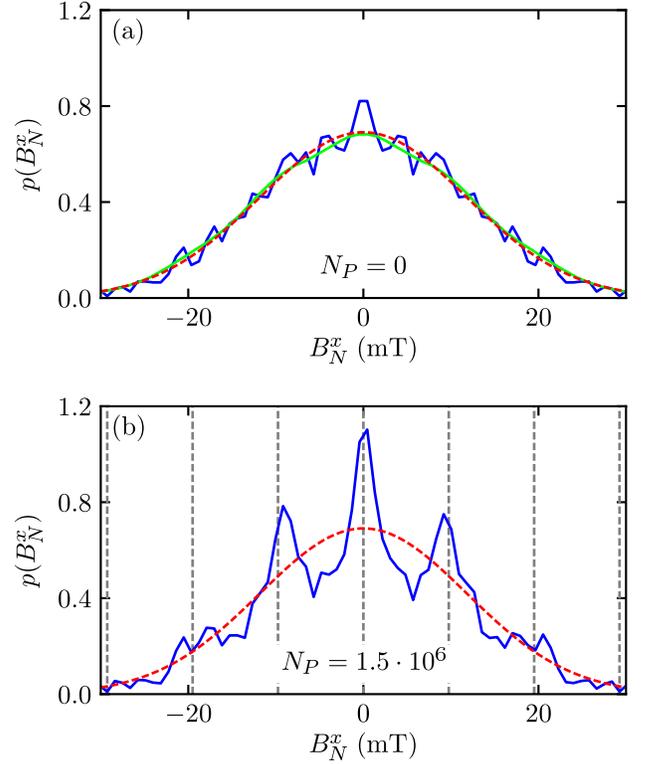}
\caption{(Color online) Overhauser field distribution $p\left( \bnx \right)$ for $\Bext = 1.95 \, \mathrm{T}$ ({i.\,e.}, $n=1$ in Eq.~\eqref{eq:nres}). The Gaussian distribution given by Eq.~\eqref{eq:gauss} is indicated by the red dashed curve. The results for a quantum mechanical system with $N=6$ and $\NC=100$ are drawn as solid lines (blue). (a) The initial distribution $p_0\left( \bnx \right)$ before the first pulse. We added  $p_0\left( \bnx \right)$ obtained for $\NC=10^5$ in green for comparison as well. (b) $p\left( \bnx \right)$ after $1.5$ million pump pulses. Overhauser fields, that correspond to an integer number of electron Larmor revolutions within $\TR$, are indicated by the grey dashed vertical lines.}
\label{fig:ohf}
\end{center}
\end{figure}

\begin{figure*}[htbp]
\begin{center}
\includegraphics[scale=1]{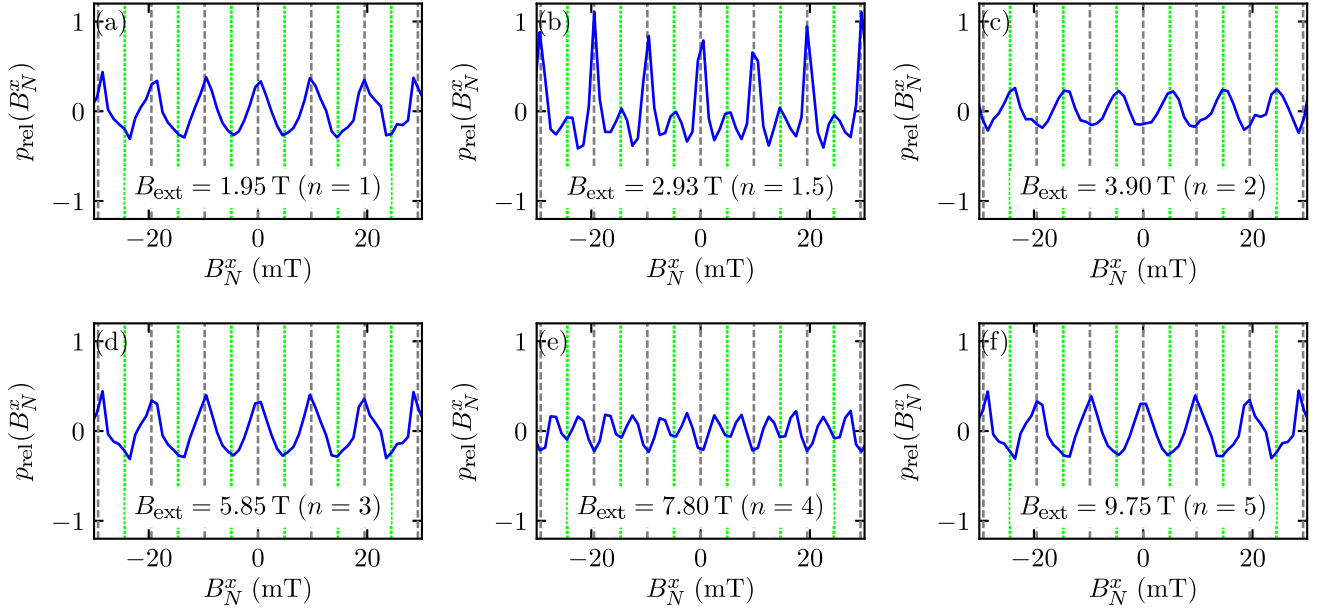}
\caption{(Color online) Relative Overhauser field distribution $p_\mathrm{rel}\left( \bnx \right)$ for distinct external magnetic fields $\Bext$. Overhauser fields, that correspond to an integer/ a half-integer number of electron Larmor revolutions, are indicated by grey dashed / green dotted vertical lines, respectively. The number $\NP$ of pump pulses is in the range $1.5 \cdot 10^6 \leq \NP \leq 20 \cdot 10^6$.}
\label{fig:relohf}
\end{center}
\end{figure*}

During the pulse sequence, the nuclei align along the external magnetic field axis in such a way that the electron spin performs a certain number of revolutions during the Larmor precession between two successive pump pulses.
The number $m$ of electron spin revolutions during $\TR$ is determined by the combination of external magnetic field and Overhauser field
\begin{align}
m = \frac{\ge \muB \TR}{2 \pi \hbar} \left( \bnx + \Bext \right) \, .
\end{align}
We adjust $\Bext$ such that for a zero Overhauser field the electron spin performs an integer number of revolutions
\begin{align}
\Bext = m' \cdot \frac{2 \pi \hbar}{\ge \muB \TR}
\end{align}
with $m' \in \mathbb{Z}$.
Since we start in the high temperature limit with an initial density operator $\rho \propto \hat 1$, the initial Overhauser field distribution $p_0\left( \bnx \right)$ is approximately a Gaussian due to the central limit theorem.
For constant $A_k$, the distribution would be binomial.
The numerical result for the initial distribution is shown in Fig.~\ref{fig:ohf} (a) for $N=6$ and $\NC=100$ (blue curve).
The finite size noise arises from the mismatch between the discrete but random eigenvalue spectrum of the operator $B_{N}^x$:
The larger $\NC$, the more continuous the eigenvalue spectrum, the smoother the distribution will be even for small $N$.
The distribution $p_0\left( \bnx \right)$ for $N \rightarrow \infty$ approaches a Gaussian~\cite{MerkulovEfrosRosen}
\begin{align}
p \left( \bnx \right) = \sqrt{\frac{3}{2 \pi}} \exp \left( - \frac{3}{2} \cdot \left( \frac{\ge \muB T^*}{\hbar} \bnx \right)^2 \right)
\label{eq:gauss}
\end{align}
in accordance with the central limit theorem and is added to Fig.~\ref{fig:ohf} for comparison (the red dashed curve).

During the pump sequence, $p\left( \bnx \right)$ evolves into a peaked structure.
The Overhauser field distribution calculated after $\NP=1.5\cdot 10^6$ pulses is shown in Fig.~\ref{fig:ohf}~(b) for a fixed external magnetic field, $\Bext = 1.95 \, \mathrm{T}$.
The maxima coincide with an integer number of electron spin revolutions during $\TR$ (marked by grey dashed lines).

In order to reduce the finite size noise, we define a relative Overhauser field distribution
\begin{align}
p_\mathrm{rel}\left( \bnx \right) = \frac{p\left( \bnx \right) - p_0\left( \bnx \right)}{p_0\left( \bnx \right)}
\end{align}
accounting for the normalized difference of $p\left( \bnx \right)$ to the initial distribution $p_0\left( \bnx \right)$ \cite{BeugelingUhrigAnders2016}.

The relative distributions of the Overhauser field in Fig.~\ref{fig:relohf}~(a)-(f) reveal a dependency of the peak position on the external magnetic field.
In Ref.~\cite{BeugelingUhrigAnders2017}, a nuclear resonance condition
\begin{align}
\Bext = n \cdot \frac{\pi \hbar}{2 \TR \gN \muN} \approx n \cdot 1.95 \, \mathrm{T}
\label{eq:nres}
\end{align}
attributed to the nuclear Zeeman term~\eqref{eq:nz} was proposed, where $n$ counts the number of quarter turns of the nuclear spins within $\TR$.
An even $n$ favors a half-integer number $m$ of electron spin revolutions (half-integer resonance) while an odd $n$ favors an integer number $m$ within $\TR$ (integer resonance).
The relative Overhauser field distribution for integer $n$ defined in Eq.~\eqref{eq:nres} displays peaks \cite{BeugelingUhrigAnders2017} at either the grey dashed lines (half-integer $m$ for even $n$) or the green dotted lines (integer $m$ for odd $n$).
For non-integer $n$, the relative Overhauser field distribution has peaks at both positions, the grey dashed and the green dotted ones (see Fig.~\ref{fig:relohf}~(b)).
For higher magnetic field, such as $7.80 \, \mathrm{T}$ ($n=4$), that have not been treated in Ref.~\cite{BeugelingUhrigAnders2017}, we observe effects additional to Eq.~\eqref{eq:nres}.
Here, we would have expected half-integer peaks, but peaks between the integer and the half-integer positions occur.

\subsection{Analysis of the steady-state revival amplitude}
\label{sec:revampohf}

So far, we presented the results of a very expensive numerical calculation to iteratively solve 
the combination of a short laser pulse that has been treated as instantaneous and the propagation of an open quantum system
between two pulses repetitively up to 20 million times.
Now we present a simplified analysis that reveals the essential connection
between the revival amplitude and the Overhauser field distribution.

We make use of the fact that at larger magnetic fields, (i) the effect of the Knight field 
defined  in Eq.\ \eqref{eq:knight-field}
is weak compared to the nuclear Zeeman term and
(ii) the collective rotation of all nuclear spins around the external field direction only very weakly changes the
transversal component of the total effective magnetic field
that the central spin is observing.
The major additional contributions
to the external magnetic field arise from the $x$-component of the
Overhauser field that is quasi-static on the time scale of $\TR$.
While the quantum mechanical calculation presented above accounts for the full dynamics of the problem,
we explore a frozen Overhauser field approximation in this section, assuming a quasi-static Overhauser field distribution.
This is justified analytically, by inspecting the magnitude of the individual $A_k$ entering the Hamiltonian or by
the explicated demonstration of  a very slow change of the revival amplitude with the number of pulses as presented in Fig.\
\ref{fig:revampnp}.

To derive the relation between Overhauser field distribution and revival amplitude, we start by treating a single configuration $K$ of nuclear spins (in configuration $j$).
For this purpose, we consider the spin component $\SzTR$, which matches the 
revival amplitude $\SpTR$ in Fig.~\ref{fig:revampb} for $\Bext \geq 2 \, \rm T$ almost perfectly.
Since a pump pulse does not act on the 
the nuclear spins, we can relate the expectation value of the spin component $\Sz_{K,j}^{b}$ before and $\Sz_{K,j}^{a}$ after the $\NP$-th pump pulse by \cite{Jaeschke2017, BeugelingUhrigAnders2017}
\begin{align}
\SzTR_{K,j}^{a} = \frac12 \left( \SzTR_{K,j}^{b} - \frac{\hbar}{2} \right) \, .
\label{eq:sxpump}
\end{align}
For the time evolution between pump pulses, we neglect the
effect of the trion decay under the assumption $\gamma \ll \omega_{e}$ and consider the nuclear spins as frozen.
Note that the nuclear spins still rotate
around the external magnetic field, but these additional components are 
small and oscillating 
compared to the total effective field in $x$-direction and only will generate a very small perturbative effect in an external field that is
two orders of magnitude larger than the Overhauser field.
This leads to the simplified relation
\begin{align}
\expect{S^z \left( \left( \NP + 1 \right) \TR \right) }_{K,j}^{b} = & \SzTR_{K,j}^{a} \notag \\
& \cdot \cos \left( \left( \omega_{e} + \omega_{K,j} \right) \TR \right) \, ,
\label{eq:sxevol}
\end{align}
where we introduced the electron Larmor frequency $\omega_{K,j} = \ge \muB B_{K,j} / \hbar$ in the Overhauser field $B_{K,j}$.
Iterating Eq.~\eqref{eq:sxpump} and Eq.~\eqref{eq:sxevol} and assuming a steady-state with constant revival amplitude $\SzTR_{K,j}^{b} = \expect{S^z \left( \left( \NP + 1 \right) \TR \right) }_{K,j}^{b}$, we obtain
\begin{align}
\SzTR_{K,j}^{b} = - \frac{\cos \left( \left( \omega_{e} + \omega_{K,j} \right) \TR \right) \, \hbar}{ 4 - 2 \cos \left( \left( \omega_{e} + \omega_{K,j} \right) \TR \right)} \, .
\label{eq:sxtrj}
\end{align}
The total revival amplitude
\begin{align}
\SzTR = \sum_{K,j} p_{K,j} \SzTR_{K,j}^{b}
\label{eq:sxtr}
\end{align}
in this approximation results from the sum over all nuclear configurations $K$ and coupling sets $j$ weighted by their probability $p_{K,j}$ introduced in Eq.~\eqref{eq:pKj}.
We make use of the fact that the electron spin dynamics is very fast
in comparison to a very slow change of nuclear spin distribution encoded in probabilities $p_{K,j}$.

Eq.~\eqref{eq:sxtr} is the central result of this section: It relates the steady-state revival amplitude obtained in a frozen 
Overhauser field approximation
and the probability $p_{K,j}$ for a specific Overhauser field configuration $K,j$ to the total revival amplitude. The quality of this
approximation relies on the separation of time scales: while the electronic steady-state is reached rather fast after only a few pulses
as demonstrated in Fig.\ \ref{fig:timeevolution}(a), the Overhauser field distribution and, therefore, the probability $p_{K,j}$
evolves very slowly on the scale of thousands of pulses -- see also Ref.\ \cite{BeugelingUhrigAnders2016,BeugelingUhrigAnders2017}.

For the calculation of $\SzTR$ according to Eq.~\eqref{eq:sxtr}, we use the weights $p_{K,j}$ obtained from the full numerical simulation.
We added the results as function of the external magnetic field into Fig.~\ref{fig:revampb} as green diamonds.
They match the amplitude of the full quantum mechanical calculation very well except for magnetic 
fields below $1\,$T, where the trion decay must be properly taken into account  
and the frozen Overhauser approximation becomes less justified.
This agreement clearly demonstrates that the revival amplitude is 
fully determined by the Overhauser field distribution $p_\mathrm{rel}\left( \bnx \right)$.
Thus, the maxima/minima of the revival amplitude coincide with odd/even 
$n$ in the resonance condition \eqref{eq:nres}, respectively.

For the continuous Gaussian distribution in Eq.~\eqref{eq:gauss}, the initial revival $\SzTR / \hbar = -0.077$ of the electronic steady-state \cite{BeugelingUhrigAnders2016}, that has been deduced in Appendix \ref{app:elss}, results directly from Eq.~\eqref{eq:sxtr}.

At the end of a very long pulse sequence, the Overhauser field distribution has a peaked structure.
We divide these peaks into two subgroups: one corresponding to the integer resonance and one for the half-integer resonance.
Assuming $\delta$-peaks for each subgroup distribution, we obtain the value $\SzTR = -1/2$ for the integer case and the value $\SzTR = 1/6$ for the half-integer case from Eq.~\eqref{eq:sxtrj}, respectively (cf. Ref.~\cite{Jaeschke2017}).
These values are independent of the resonance Larmor frequency $\omega_{K,j}$.
Thus, the weights $p_{K,j}$ do not enter the full revival amplitude in Eq.~\eqref{eq:sxtr}.
Since the steady-state amplitudes have opposite signs for the different resonance conditions a destructive interference between these two subsets is found \cite{Jaeschke2017} and the final value depends on the ratio between the fractional weights of these parts.
Compared to the initial value $\SzTR / \hbar = -0.077$ ($\NP \approx 10$) in the electronic steady-state, the electron spin component $|\SzTR|$ either increases (integer case) or decreases (half-integer case).
As the peaks in our numerical calculation of the Overhauser field retain a finite width, the revival amplitude results from a superposition of the contributions from both resonances explaining the evolution of $\SpTR$ as depicted in Fig.\ \ref{fig:revampnp}.

By means of the resonance condition for the Overhauser field distribution, we are now able to understand why the behavior of the revival amplitude is more complex in the experiment (cf. Fig.~\ref{fig:exp_magn_field}) than presented in Fig.~\ref{fig:revampb}.
We have simplified our theoretical model to a single type of nuclei with a single average $g$-factor, whereas in real samples the $g$-factor differs between the elements In, Ga and As as well as between the respective isotopes of an element. 
We observed the magnetic field dependency of the revival amplitude stemming from the resonance condition \eqref{eq:nres} for different values of the nuclear $g$-factor $\gN$ determining the number of nuclear spin revolutions in the time $T_R$ (not shown here).
For increasing $\gN$, the minimum of the revival amplitude shifts to lower magnetic fields.
Results presented in Ref.~\cite{BeugelingUhrigAnders2017} indicate that each type of nucleus leads to a separate resonance condition in the form of Eq.~\eqref{eq:nres}.
The different kinds of peaks in the Overhauser field distribution are more or less pronounced depending on the external magnetic field and which resonances of the various nuclear species are closest.
As a result, the behavior of the revival amplitude is expected to become more complex when involving several types of nuclei.
Since the individual $g$-factors of most nuclei induce a minimum of revival amplitude between $3.7 \, \rm T$ and $5.2 \, \rm T$ according to Eq.~\eqref{eq:nres} ($n=2$), the combined behavior results in a minimum at around $4 \, \rm T$ for both samples in Fig.~\ref{fig:exp_magn_field}.
Additional non-monotonic behavior distinguishing the samples can be attributed to the different concentration of the nuclear species in the QDs, {e.\,g.}, due to the different thermal annealing of the samples.
At higher external magnetic fields, the resonance condition for the different nuclear species disperses more strongly leading to a decrease of the total revival amplitude for both samples in Fig.~\ref{fig:exp_magn_field}.

\section{Classical approach to the quantum dynamics of periodically driven QDs }
\label{sec:classic-pulses}

A non-monotonic dependence of the revival amplitude on the external magnetic field is also obtained in an advanced classical approach simulating the quantum dynamics.
In this approach, the central electron spin and the nuclear spin bath are treated as classical vectors, but the average is taken over Gaussian distributed initial conditions which mimics the quantum mechanical dynamics \cite{erlin04,Chen2007,alhas06,polko10,Stanek2014}.
The details of the approach are developed and analyzed in detail in Refs.\ \cite{Fauseweh2017,Schering2018}.

We calculate the full time evolution of the classical equations of motion of the CSM \eqref{eq:CSM} for generically distributed dimensionless hyperfine couplings 
$\{A_k\}$
\begin{align}
A_k = \mathcal{C} \mathrm{e}^{-k \zeta}\, ,
\end{align}
where $\zeta$ replaces the parameter $\gamma$ in Ref.~\cite{Schering2018}.
In the numerics, $\mathcal{C}$ is chosen such that $A_\mathrm{Q} := \sqrt{\sum_k A_k^2}$ is set to unity, {i.\,e.}, all energies are measured in units of $A_\mathrm{Q}$.
In order to enable a quantitative comparison to the experiment, we set $\hbar/A_\mathrm{Q}=0.79 \,$ns, which implies that for bath spins $I=3/2$ the characteristic time reads $T^*=\hbar/(A_\mathrm{Q}\sqrt{I(I+1)})=0.41 \,$ns according to Eq.\ \eqref{eq:tstar_noensemble}, in good agreement with the experiment \cite{GreilichScience2006,GreilichScience2007,PhysRevLett.96.227401} (cf. Sec.\ \ref{sec:csm}).

The average over $10^4 - 10^5$ random initial configurations is used to approximate the quantum mechanical behavior of a single quantum dot \cite{Chen2007,Stanek2014}.
The initial values of each configuration are drawn from a Gaussian distribution with vanishing average value and a variance reflecting the spin length, {i.\,e.}, $1/4$ for each component of the central electron spin and $5/4$ for each component of a nuclear spin with spin $I = 3/2$.

The full time evolution of the Overhauser field is simulated efficiently by the spectral density approach developed in Ref.\ \cite{Fauseweh2017}.
It allows us to consider an infinite spin bath, while the number of \emph{effectively} coupled spins is finite and given by $N_\mathrm{eff} \approx 2/\zeta$ \cite{Fauseweh2017,Roehrig2018,Schering2018}.
In our calculations, we use between $44$ and $74$ auxiliary vectors, where the exact number depends on $\NP$ (cf. Refs.\ \cite{Schering2018,Fauseweh2017}), to represent bath sizes of up to $N_{\rm eff}=667$.
The Zeeman effect of the magnetic field applied to the central spin is taken into account by adding a term $h S^x$ with $h=g_{e}\mu_{B} B_\mathrm{ext}/\hbar$, while the Zeeman effect of the nuclear spins is reduced by the factor $z\ll1$ according to $h\to zh$.
The value $z=1/800$ represents a good estimate \cite{BeugelingUhrigAnders2017} as discussed in Sec.\ \ref{sec:csm}.
Note that we are considering a single quantum dot here, not an ensemble.
But the extension to an ensemble of QDs is straightforward.

The quantum mechanical description of the pump pulses is involved as is evident from
the above discussion.
In the approximating classical simulation, we pursue two aims.
On the one hand, we aim at a transparent description in the classical approach.
On the other hand, it should mimic the quantum mechanical properties best.
In previous work \cite{Schering2018}, we found that the following assumption leads to convincing results.
In particular, it leads to non-monotonic revival amplitudes.

In our pulse description, the pulse affects the vector of the central spin instantaneously.
Independent of the direction prior to the pulse, right after the pulse the vector of the central spin becomes
\begin{align}
	\vec{S} \rightarrow \begin{pmatrix} X \\ Y \\ 1/2 \end{pmatrix}\,. 
	\label{eq:classical_pulse}
\end{align}
This means that we assume the pulse to be perfect in the sense that it produces maximum alignment along the $z$-axis.
The values of $X$ and $Y$ are chosen randomly for each pulse from a Gaussian distribution with vanishing mean value $0$ and variance $1/4$.
This randomness is introduced to respect Heisenberg's uncertainty relation for the electron spin which forbids a perfect alignment.
Additionally, it ensures that the expectation value for the spin length $\langle \vec{S}^2\rangle$ takes the correct value of $3/4$.
To consider this sort of classical pulse mimicking quantum mechanics is motivated by viewing the pulse as a quantum mechanical measurement with a definite outcome for the $z$-component.
In Ref.\ \cite{Schering2018}, this type of pulse was denoted as pulse model II.
It represents an extension of the pulses studied in previous works \cite{PetrovYakovlev2012, Schering2018}. 

\begin{figure}[t]
	\begin{center}
		\includegraphics[width=\columnwidth]{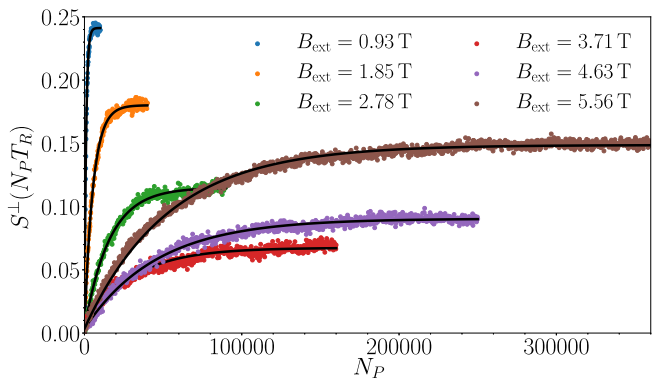}
		\caption{(Color online) Evolution of the revival amplitude 
		$S^\perp(N_{P} T_{R})$ as function of the  pulse number $N_{P}$
		for various external magnetic fields $B_\mathrm{ext}$, averaged over $25200$ 
		random initial configurations with $N_\mathrm{eff} = 200$. The solid black lines show the 			fits \eqref{eq:fit_classical} yielding the saturated value $S_\mathrm{lim}$ of the revival 			amplitude.}
		\label{fig:classical_Slim_fit}
	\end{center}
\end{figure}

The used values of the parameters, {e.\,g.}, $T^*$, differ slightly from those used in Sec.~\ref{sec:qmapproach}, but still correspond to the values typical for (In,Ga)As/GaAs quantum dots as measured in Sec.~\ref{sec:experiment}.
Hence, the results can be compared at least qualitatively. 

Simulating up to $10^6$ pulses, we are able to reliably extrapolate a value for the saturated revival amplitude $S_\mathrm{lim}$.
The explicit value is calculated by fitting the function
\begin{align}
	S^\perp(N_{P} T_{R}) = S_\mathrm{lim,0} \left(1 - \mathrm{e}^{-N_{P} T_{R}/\tau} \right) + S_0
	\label{eq:fit_classical}
\end{align}
to the data.
Eventually, the revival amplitude is given by $S_\mathrm{lim} = S_\mathrm{lim,0} + S_0$.
This analysis is carried out for various external magnetic fields up to $10\,\mathrm{T}$ for two different effective bath sizes $N_\mathrm{eff} \approx 200$ ($\zeta=0.01$) and $667$ ($\zeta=0.003$).
An illustration of the fit procedure for various external magnetic fields is depicted in Fig.\ \ref{fig:classical_Slim_fit}.

The time required to approach the saturation value scales linearly in the inverse size of the bath $\propto \zeta \propto 1/N_\mathrm{eff}$ and quadratically in the magnetic field $\propto B_\mathrm{ext}^2$ as analyzed in Ref.\ \cite{Schering2018}.
Hence, the simulations become very tedious for large magnetic fields and large bath sizes.
Thus, we have to restrict ourselves to moderate bath sizes in this study.
But they still exceed the bath sizes which can be addressed quantum mechanically
by two orders of magnitude so that they yield complementary information.

\begin{figure}[t]
	\begin{center}
		\includegraphics[width=\columnwidth]{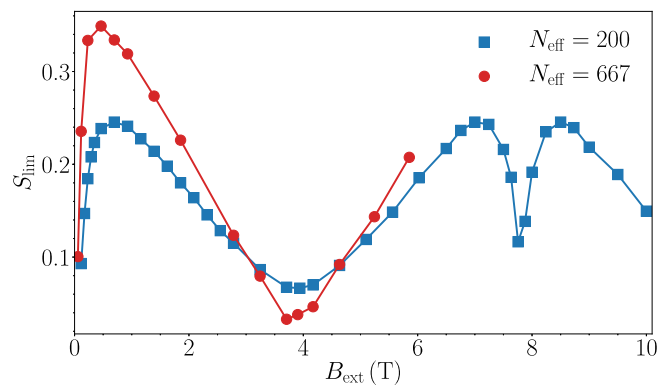}
		\caption{(Color online) Dependence of the saturated revival amplitude 
		$S_\mathrm{lim}$ on the external magnetic field. The saturated revival amplitudes 
		are determined by fitting Eq.~\eqref{eq:fit_classical} to the data from the 
		classical simulations. The lines are guides to the eye.}
		\label{fig:classical_Slim_B}
	\end{center}
\end{figure}

The results are compiled in Fig.\ \ref{fig:classical_Slim_B}.
The non-monotonic dependence of $S_\mathrm{lim}$ on the external magnetic field $B_\mathrm{ext}$ shows a pronounced minimum at around $4\,\mathrm{T}$ similar to what is found in Fig.\ \ref{fig:revampb} for the quantum mechanical approach although the details are different.
A less pronounced minimum occurs at around $8$\,T which is much narrower than what is found quantum mechanically.
Additionally, there is a maximum slightly below $1\,\mathrm{T}$.
Such a maximum is also found in the quantum mechanical approach, but shifted to larger magnetic fields, which may result from the difference in bath sizes and from the difference between the full quantum mechanical dynamics and the classical simulation.

The comparison to the experimental data in Fig.\ \ref{fig:exp_magn_field} also reveals strong similarities such as the pronounced minimum at about $4\, \rm T$ and weaker structures at around $8\, \rm T$.
The minimum at $8$\,T is very narrow and requires many data points to be resolved correctly.
Note also the similarity to the experimental data for the revival amplitude published in Fig.\ 20 in Ref.\ \cite{Jaeschke2017}.
But the position of the maximum to the left of the minimum differs because there are additional experimental features.
We presume that they result from the different species of nuclei present in the samples as discussed in Sec.~\ref{sec:revampohf}.
Concomitantly, there are five different $g_{N}$-factors, which one should consider while our theoretical treatments deal with one average $g_{N}$-factor only.
In addition, the classical simulation does not treat an ensemble of QDs, {i.\,e.}, the effects of a spread in $T^*$ and in the electronic $g$-factor is not yet included.

We emphasize that in the classical simulation, the build-up of the revival amplitude is solely due to the frequency focusing of the nuclei, {i.\,e.}, of the formation of a comb-like structure in the distribution of the Overhauser field.
The contribution from the electronic steady-state condition is not included.
Fig.~\ref{fig:classical_Overhauser} shows the almost stationary distribution of the $x$-component of the Overhauser field for two external magnetic fields $B_\mathrm{ext} = 0.93\,\mathrm{T}$ and $3.71\,\mathrm{T}$.
As an aside, we note that it is not the $x$-component of the total magnetic field $\vec{B}$, external and Overhauser, which matters \cite{Schering2018}, but its length $|\vec{B}|$.
The first magnetic field corresponds to the blue curve for the revival amplitude in Fig.\ \ref{fig:classical_Slim_fit}, the second field to the red curve.
Both distributions show a comb-like structure of nuclear focusing with peaks corresponding to the \emph{integer} resonance condition, {i.\,e.}, for an integer number of electron spin revolutions within the interval $\TR$ between two pulses.
But the width and concomitantly the height of the peaks differ substantially.
This explains the much smaller revival amplitude for the magnetic field close to $4 \, \rm T$ and $8 \, \rm T$.
In general, we find that a larger value of the revival amplitude corresponds to sharper peaks.
We do not observe additional peaks at the Overhauser fields corresponding to half-integer resonances.

\begin{figure}[t]
	\begin{center}
		\includegraphics[angle=90,width=\columnwidth]{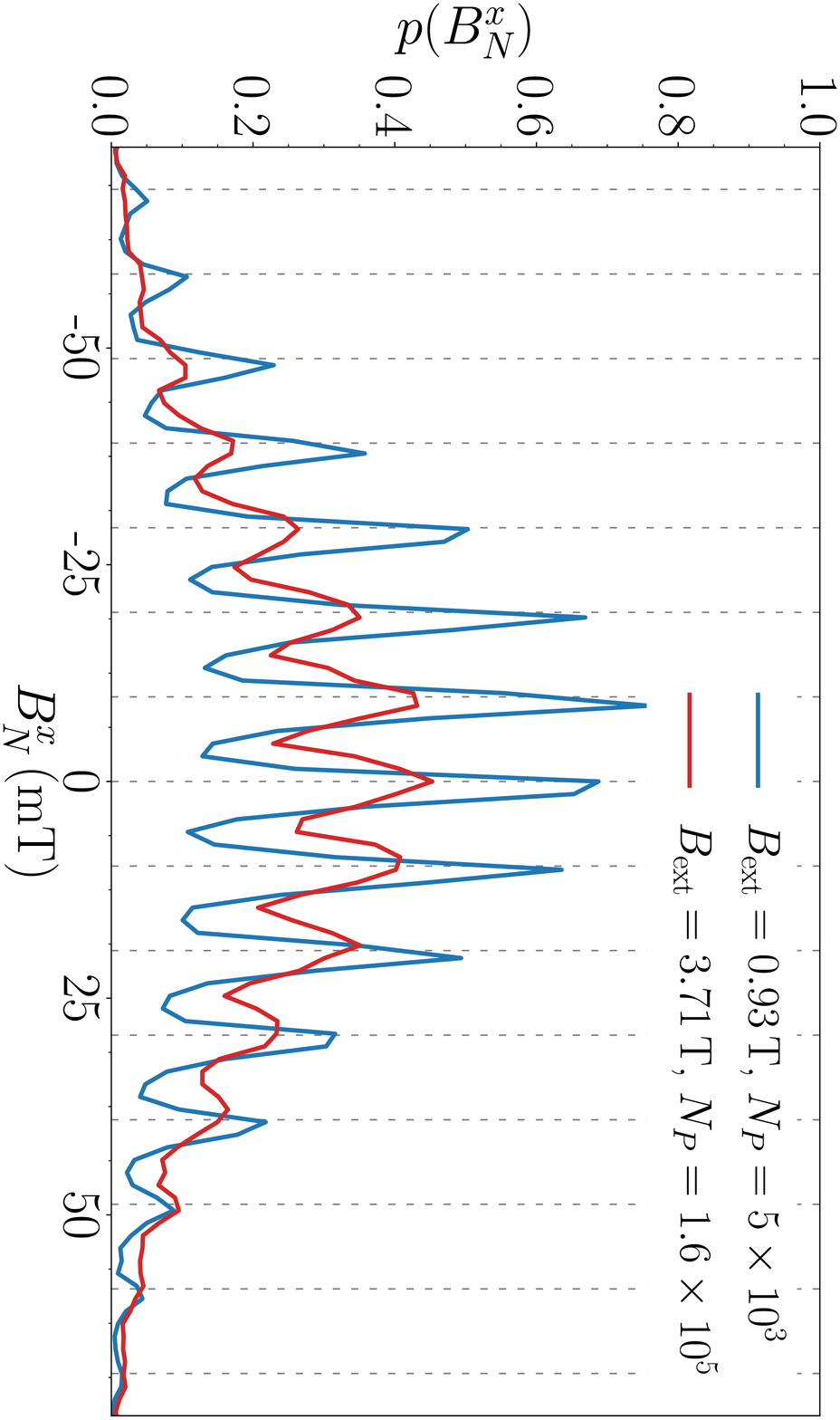}
		\caption{(Color online) Distribution of the Overhauser field $x$-component obtained from the classical simulations. The calculations were performed for an ensemble of $25200$ random initial configurations with $N_\mathrm{eff} = 200$. The vertical dashed lines indicate the integer resonance condition.}
		\label{fig:classical_Overhauser}
	\end{center}
\end{figure}

The build-up of nuclear frequency focusing in the classical simulations has been studied in detail in Ref.\ \cite{Schering2018}. 
For the pulse model \eqref{eq:classical_pulse}, it was found that the build-up rate scales approximately with $1/N_\mathrm{eff}$.
No perfect scaling was found so that there remains a dependence on the bath size;
this is also manifest in Fig.\ \ref{fig:classical_Slim_B} where the long-time minima and maxima are more pronounced for larger $N_\mathrm{eff}$.
For the dependence on the external magnetic field, a non-monotonic behavior was found \cite{Schering2018}.
However, the overall time required to reach a stationary Overhauser field distribution, and hence a saturated revival amplitude, scales approximately with $B_\mathrm{ext}^{2}$ (cf.\ Fig.\ \ref{fig:classical_Slim_fit}).

What is the reason in the classical simulations for the non-monotonic dependence on the external magnetic field shown in Fig.\ \ref{fig:classical_Slim_B}?
It does capture the interplay of electronic and nuclear precessions.
An important additional clue is obtained from setting $X = Y = 0$ in each classical pulse \eqref{eq:classical_pulse}, {i.\,e.}, from neglecting the uncertainty in the spin orientation.
Then, the revival amplitude saturates at $S_\mathrm{lim} = 1/2$, totally independent of the value of the applied external magnetic field (cf.\ Ref.\ \cite{Schering2018}).
The distribution of the $x$-component of the Overhauser field displays very sharp peaks at positions corresponding to the integer resonance condition leading to a perfect refocusing of the electron spin precession before each next pulse, {i.\,e.}, to a maximum revival amplitude.
Hence, it is indeed the quantum uncertainty, mimicked by the randomness of $X$ and $Y$ in the classical simulations, which is decisive for the \emph{finite} peak widths shown in Fig.\ \ref{fig:classical_Overhauser} which imply the reduced revival amplitude and eventually the non-monotonic behavior depicted in Fig.\ \ref{fig:classical_Slim_B}.

An additional piece of information, in which way the randomness in $X$ and $Y$ acts against perfect nuclear focusing, results from the following observation for a magnetic field around 4 T.
Including only the fluctuations in the $y$-component results in half-integer resonances while including only the fluctuations in the $x$-component results in integer resonances.
Hence, they act against each other and the reduced nuclear focusing is an effect of destructive interference.
Clearly, it will be attractive to clarify this issue further by analytical considerations.

\section{Non-instantaneous pump pulses in the quantum mechanical approach}
\label{sec:pulses}

So far, we only took into account instantaneous $\pi$-pulses that resonantly excite the trion in the quantum mechanical approach and affect the electron spin in the classical simulation.
Experimental pump pulses, however, have a finite duration of a few picoseconds \cite{GreilichScience2006,GreilichScience2007,PhysRevLett.96.227401,GREILICH20091466}.
A deviation from a perfect resonance condition due to the electronic Zeeman energy as well as the spin precession during the pulses might affect the steady-state revival amplitude at large external magnetic fields.
Furthermore, an extension to arbitrary pulse shapes will open a new door for more complex pulse sequences in the future.

\begin{figure}[b]
\begin{center}
\includegraphics[scale=1]{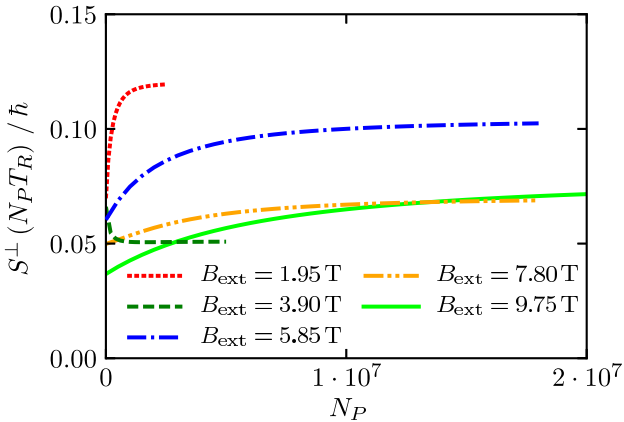}
\caption{(Color online) Evolution of the electron spin revival amplitude with the pulse number $\NP$ for Gaussian pump pulses in the quantum mechanical approach. Various colors show the development for different external magnetic fields $\Bext$.}
\label{fig:revampnpgauss}
\end{center}
\end{figure}

As a first step for more realistic pulses, we consider Gaussian pump pulses in the quantum mechanical approach.
Thus, we need to replace the unitary pulse operator $U_{P}$ by a new operator $U_{P}'$.
This operator $U_{P}'$ is obtained by integrating the equation of motion for the unitary time evolution during the pulse duration.
For this purpose, we use the light-matter Hamiltonian in rotating wave approximation \cite{Carmichael}
\begin{align}
H_{\rm L} \left( t \right) = &f(t) \mathrm{e}^{-\im \omega_{L} t / \hbar} \ket{\mathrm{T}} \bra{\up}_z + \mathrm{H. \, c.} \, ,
\end{align}
where $\omega_{L}$ denotes the laser frequency.
The Gaussian pulse shape is included in the (complex) envelope function $f(t)$.
During the pump pulse, the total Hamiltonian is given by $H(t) = H_{\rm L} (t) + \Hcsm + \epsilon \ket{\mathrm{T}} \bra{\mathrm{T}}$.
The trion decay is neglected, as the decay rate $\gamma = 10 \, \mathrm{ns}^{-1}$ is slow compared to the duration $T_{P}$ of the pulse.

\begin{figure}[b]
\begin{center}
\includegraphics[scale=1]{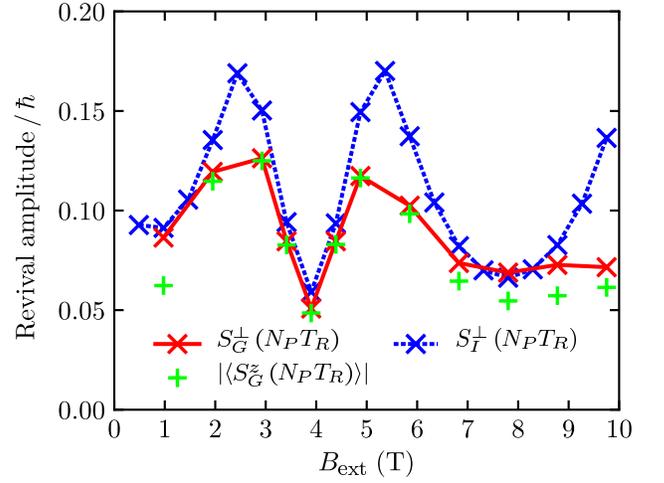}
\caption{(Color online) Magnetic field dependency of the electron spin revival amplitude calculated by the quantum mechanical approach with Gaussian pump pulses. The revival amplitude $S_{G}^\perp \left( \NP \TR \right)$ and the spin component $\left| \left< S_{G}^z \left( \NP \TR \right) \right> \right|$ are taken after a number of pump pulses $2.5 \cdot 10^6 \leq \NP \leq 20 \cdot 10^6$ large enough such that they have converged. 
The exact value of $\NP$ depends on the magnetic field. 
For comparison, we added the revival amplitude $S_{I}^\perp \left( \NP \TR \right)$ with instantaneous pump pulses taken from Fig.~\ref{fig:revampb}.}
\label{fig:revampbgauss}
\end{center}
\end{figure}

\begin{figure*}[t]
\begin{center}
\includegraphics[scale=1]{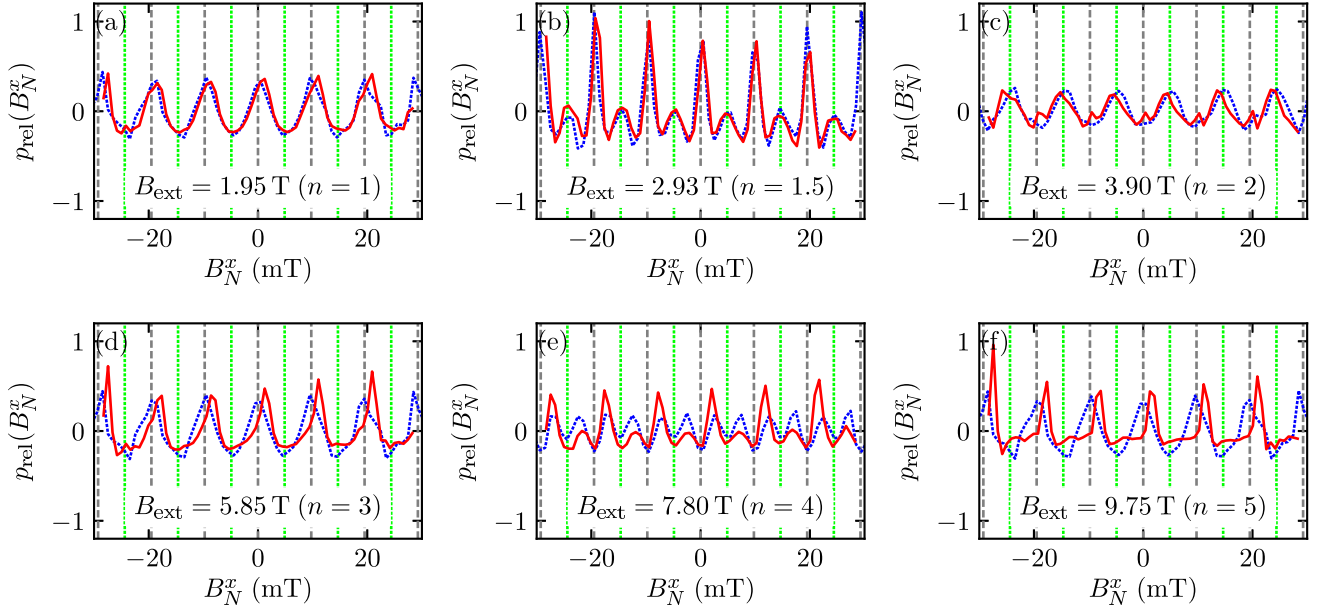}
\caption{(Color online) Relative Overhauser field distribution $p_\mathrm{rel}\left( \bnx \right)$ for various external magnetic fields $\Bext$. The considered pulse sequence consists of Gaussian pump pulses (red solid lines). 
Overhauser fields, that correspond to an integer/ a half-integer number of electron spin revolutions during $\TR$, are indicated by grey dashed / green dotted vertical lines respectively. The number $\NP$ of pump pulses is in the range $2.5 \cdot 10^6 \leq \NP \leq 20 \cdot 10^6$. 
For comparison we added the results for instantaneous pump pulses taken from Fig.~\ref{fig:relohf} as blue dotted lines.}
\label{fig:relohfgauss}
\end{center}
\end{figure*}

First, we transform into the frame rotating with the laser frequency $\omega_{L}$ and eliminate the fast oscillation with $\omega_{L}$ in the Hamiltonian.
Introducing the detuning $\delta = \omega_{L} - \epsilon$, we obtain the transformed Hamiltonian
\begin{align}
H'(t) = &\mathrm{e}^{\im \omega_{L} \ket{\mathrm{T}} \bra{\mathrm{T}} t / \hbar} \left( H(t) - \omega_{L} \ket{\rm T} \bra{\rm T} \right) \mathrm{e}^{- \im \omega_{L} \ket{\mathrm{T}} \bra{\mathrm{T}} t / \hbar} \notag \\
= &f(t) \ket{\mathrm{T}}\bra{\up}_z + f^*(t) \ket{\up}_z \bra{\mathrm{T}} \notag \\
&+ \Hcsm - \delta \ket{\mathrm{T}} \bra{\mathrm{T}} \, .
\end{align}
Second, we discretize the Hamiltonian  $H'(t)$ in small time steps, defining intervals for which $f(t_n)$ can be considered approximately as constant.
In our numerics, we typically divide a single laser pulse in $1000$ time steps so that $\Delta t\approx 22 \, {\rm fs}$ for a total pulse duration $T_{P} \approx 22 \, {\rm ps}$.
The unitary time evolution is approximated by operators
\begin{align}
U(t_n) = \mathrm{e}^{- \im H'(t_n) \Delta t / \hbar}
\end{align} 
and their Hermitian conjugates $U^\dag (t_n)$, where $\Delta t = t_{n+1} - t_{n}$ is the step width in time.
Neglecting the Trotter error, which vanishes for $\Delta t\to 0$, the unitary transformation is given by the product of all individual transformations
\begin{eqnarray}
U_{P}'  &=&   e^{-\im \omega_{L} \ket{\rm T} \bra{\rm T} T_{P} / \hbar}\prod_{n} U(t_n)
\nonumber \\
& =& e^{-\im \omega_{L} \ket{\rm T} \bra{\rm T} T_{P} / \hbar}U(T_{P}) ... U(t_2) U(t_1) \, .
\end{eqnarray}
Note that the additional exponential factor accounts for the back transformation from the rotating frame.
The transformation of the density matrix into the rotating frame is omitted.

The pulse action is described by
\begin{align}
\rho \left( T_{P} \right) = & U_{P}' \rho (0) U_{P}'^\dag \, ,
\label{eq:rhoTP}
\end{align}
where $\rho (0)$ and $\rho \left( T_{P} \right)$ are the density operator before and after the pulse respectively.
Since the unitary transformation is obtained initially and stored as one unitary complex matrix, modified pulses just come at the expense of two additional complex
matrix multiplications in the numerical implementation.

After each pulse, the time evolution is again calculated using the Lindblad equation discussed in Sec.~\ref{sec:lindblad} with $\rho \left( T_{P} \right) $ obtained via  Eq.~\eqref{eq:rhoTP} as input.
Since the pump pulse now has a finite duration $T_{P}$, we evaluate the time evolution via Lindblad equation for a reduced duration $\TR - T_{P}$.

For the calculations presented below, we choose a Gaussian pulse shape with $f(t) = f^*(t)$, whose iterated area corresponds to a $\pi$-pulse.
The full width at half maximum (FWHM) is adjusted to $6 \, \rm ps$.
This width is slightly larger than in the experiments, but renders possible effects on the spin dynamics more visible.
The duration of the pulse is set to $T_{P} \approx 22 \, {\rm ps}$, within which we consider the part of the pulse up to which the envelope $f(t)$ has decayed to a hundredth of its maximum.
For the laser frequency, we restrict ourselves to $\omega_{L} = \epsilon$, such that the trion is resonantly excited without detuning ($\delta = 0$), and leave the investigation of the influence of the detuning in a quantum dot ensemble to future studies.

We use the same parameters as before, but replace the instantaneous $\pi$-pulses by Gaussian shaped pulses with a finite width.
The pulse number dependent revival amplitude for such Gaussian pulses is shown in Fig.~\ref{fig:revampnpgauss} for the same external magnetic field values as in Fig.\ \ref{fig:revampnp}.
In comparison with the result for instantaneous pump pulses in Fig.~\ref{fig:revampnp}, a slower rate of change is observed.
Therefore, a larger number of pump pulses is required to reach a converged steady-state revival amplitude.
Especially for higher magnetic fields, the pump pulses become less efficient, as the electron spin precesses during the pulse duration. 

To investigate the influence of the Gaussian pump pulses on the magnetic field dependency, the converged revival amplitude $\SpTR$ is again plotted as function of $\Bext$.
The result in Fig.~\ref{fig:revampbgauss} (red curve) shows some difference to the data
for instantaneous pump pulses taken from Fig.~\ref{fig:revampb} 
which we added for comparison (blue curve), even though the overall qualitative behavior remains  the same.
There are still two maxima and two minima respectively in the magnetic field range up to $10 \, \mathrm{T}$, but the maximum amplitude has decreased.
Besides, the amplitude of the second maximum is smaller than the amplitude of the first maximum.
Note that the revival amplitude for the data point at $\Bext = 9.75 \, \mathrm{T}$ is not completely converged (see Fig.~\ref{fig:revampnpgauss} green curve) and therefore the minimum at about $8\, \rm T$ is not very pronounced.
However, we again observe that the revival is weaker for higher external magnetic fields.
This behavior matches the overall decrease of the revival amplitude with the external magnetic field in the experimental data in Fig.~\ref{fig:exp_magn_field}. 
Thus, the finite pulse duration is another aspect which has to be included for a realistic description of the experiments.

We augment the analysis by adding the spin component $|\SzTR|$ as green crosses to Fig.~\ref{fig:revampbgauss}.
While the spin polarization in $z$-direction agrees well with the revival amplitude $\SpTR$ in the interval $2 \, \mathrm{T} \leq \Bext \leq 6 \, \mathrm{T}$, significant deviations are found for $\Bext < 2 \, \mathrm{T}$ as well as for $\Bext > 6 \,\mathrm{T}$.
In these regions, the spin component $\SyTR$ does not vanish.

For further investigation of the Gaussian pump pulses, we inspect the relative Overhauser field distribution in Fig.~\ref{fig:relohfgauss} 
(red solid lines).
Here, we present the distributions for the same external magnetic fields as in Fig.~\ref{fig:relohf} for the instantaneous laser pulses.
The previous results for instantaneous pump pulses are added to Fig.~\ref{fig:relohfgauss} for comparison as blue dashed lines.
For external magnetic fields up to $2.93 \, \mathrm{T}$, we do not observe significant differences in the distributions for the two types of pump pulses.
However, for the first external magnetic field with even $n$ in Eq.~\eqref{eq:nres} ($\Bext = 3.90 \, \mathrm{T}$), for which we found peaks at the green dotted lines for the instantaneous pump pulses, we also find tiny peaks at the grey dashed positions for the Gaussian pump pulses.
For even higher external magnetic fields, the differences become more significant.
For $n=4$ ($\Bext = 7.80 \, \mathrm{T}$), one kind of peaks is more pronounced than the other.
The peaks for external magnetic field with odd $n$ ($\Bext = 5.85 \, \mathrm{T}$ and $\Bext = 9.75 \, \mathrm{T}$) are slightly shifted to the right from their original position at the grey dashed lines.
Therefore, the shape of the pump pulses seems to influence the resonance condition for the Overhauser field distribution and thus the electron spin revival amplitude.

\section{Summary and conclusion}
\label{sec:conclusion}

We investigated the magnetic field dependency of the revival amplitude of the electron spin polarization along the optical axis in a periodically pulsed QD ensemble.
The steady-state resonance condition leads to a significant revival directly before each pump pulse.
This has been qualitatively explained by the mode-locking of the electron spin dynamics, comprising a synchronization of the electron spin precession imposed by the periodic pumping and an enhancement by the nuclear frequency focusing that develops on a much longer time scale \cite{GREILICH20091466,GreilichScience2006,GreilichScience2007}.

The non-monotonic magnetic field dependency of the revival amplitude, however, had not been theoretically understood.
In this paper, our simulations of the CSM subject to up to $20$ million laser pulses are able to link this non-linear field dependency to the nuclear Zeeman effect. 

The quantum mechanical calculations are based on an extension of the CSM including the trion excitation due to the pump pulses.
The time evolution between two successive pump pulses including the trion decay is described by a Lindblad equation for open quantum systems that is exactly solved for each pulse interval.
Although our approach can treat arbitrary pulse shapes and durations, we focus on $\pi$-pulses in this paper.

In order to achieve pulse sequences with up to $20$ million pump pulses in our quantum mechanical approach, we restrict ourselves to a small bath of $N=6$ nuclear spins due to CPU time limitations.
Even though in real QDs an electron spin couples to the order of $10^5$ nuclear spins, it is already established that the generic spin dynamics of the CSM can already be accessed by a relatively small number of nuclei \cite{HackmannAnders2014,FroehlingAnders2017}.
We simulated a distribution of different characteristic time scales $T^*_j$ in a QD ensemble by the treatment of $\NC = 100$ configurations with distinct hyperfine coupling constants $A_{k,j}$.
The number of pump pulses required to reach a converged revival amplitude grows with increasing external magnetic field.

To support the demanding quantum mechanical computations, we also perform a classical simulation of the CSM which simulates a bath of up to $670$ effectively coupled spins \cite{Schering2018}.
This simulation is set up such that it approximates the quantum mechanical dynamics as close as possible.
But, the intermediate trion excitation and its subsequent fast decay are not built-in in the classical treatment.

Both approaches cover up to eleven orders of magnitude in times: from a single laser pulse with the duration of $2$-$10$\,ps, 
the laser repetition time of $13.2$\,ns to $20$ million pulses reaching a total simulation time of approximately $0.2$\,s.
Our key finding is that the stationary revival amplitudes exhibit a non-monotonic behavior as function of the applied external magnetic field.
There are minima of the revival amplitude at $4 \, \mathrm{T}$ and $8 \, \mathrm{T}$, which roughly match the experimental data.

In the quantum mechanical approach, the steady-state resonance conditions favor an integer or a half-integer number of electron spin revolutions between two pump pulses and eventually lead to a rearrangement of the Overhauser field distribution function similar to the one found in Refs.~\citep{BeugelingUhrigAnders2017,Jaeschke2017,Schering2018}.
The minimum of the revival amplitude is reached in case of the half-integer resonance, whereas the maximum corresponds to the integer resonance.

In the simulations, we only included a single average nuclear $g$-factor but were able to link the revival minima to the nuclear $g$-factor by variation of its value.
However, it has been indicated \cite{BeugelingUhrigAnders2017} that in real QDs the different nuclear species yield separate resonance conditions.
Since the nuclear $g$-factor is isotope-dependent the experimental response is not unique but sample-dependent.

The mechanism generating the magnetic field dependency in the classical simulations works similarly, but with one important difference.
No peaks at the Overhauser fields of the half-integer resonances occur.
Instead, the peaks corresponding to the integer resonances become broad and less pronounced for an even number of nuclear quarter turns.
Hence, the nuclear frequency focusing is little efficient and the revival amplitudes are small again due to a partial destructive interference.

We have also extended the quantum mechanical theory from instantaneous laser pulses to pulses with a finite width of $6 \, \mathrm{ps}$.
The pulses have a Gaussian shape with an area corresponding to the instantaneous $\pi$-pulses.
In this way, we take into account the possible detuning of the resonance frequency in strong magnetic field by the Zeeman effect as well as the electron spin rotation during the pulse duration.
Deviations from the instantaneous pulses occur at higher external magnetic field, 
when the electron spin rotation is non-negligible during the pulse duration.
Here, the pulse is less efficient and the formation of a revival is less pronounced.
Besides, the resonance condition for the Overhauser field is slightly shifted for higher external magnetic fields.

Even though we restricted ourselves to resonant Gaussian $\pi$-pulses, the effects of arbitrary pulse shapes as well as the detuning of laser frequency become accessible by our approach and present an interesting field for future research. 
Finite pulse lengths, detuned laser frequencies and pulse shapes, which do not correspond to $\pi$-pulses will be addressed with our approach to design specially tailored and optimized pulse trains for quantum coherent control.
Furthermore, we stress that the theoretical approaches developed and used in this work can be applied to a considerable variety of experiments on QDs subject to optical pulses.
The pulse trains need not be periodic, but could be varied to a large extent.

\begin{acknowledgments}

We are grateful for fruitful discussions on the project with A. Fischer and N. J\"aschke.
We acknowledge the supply of the quantum dot samples by D. Reuter and A. D. Wieck (Bochum).
We also acknowledge financial support by the Deutsche Forschungsgemeinschaft and the Russian Foundation of Basic Research through the transregio TRR 160 within the Projects No.\ A1, A4, A5 and A7 as well as financial support by the Ministry of Education and Science of the Russian Federation (Contract No.\ 14.Z50.31.0021, leading researcher M. Bayer). M.B. and A.G. acknowledge the support by the BMBF in the frame of the Project Q.com-H (Contract No. 16KIS0104K).
The authors gratefully acknowledge the computing time granted by the John von Neumann Institute for Computing (NIC) under Project HDO09 and provided on the supercomputer JUQUEEN at the J\"ulich Supercomputing Centre.

\end{acknowledgments}

\appendix

\section{Particular solution for the Lindblad equation}
\label{app:partsol}

To obtain a particular solution to the Lindblad Eq.~\eqref{eq:lbrhos}, we need to calculate the operators $\tilde{\chi}_0$, $\tilde{\chi}_+$  and $\tilde{\chi}_-$ in the ansatz~\eqref{eq:rhonh}.
For this purpose, we insert Eq.~\eqref{eq:rhonh} into Eq.~\eqref{eq:lbrhos}.
Separating the terms according to the three different exponents in the exponential functions, yields the conditions ($\alpha \in \{ 0 , + , - \}$)
\begin{align}
\left( \im \omega_{N} \left( \delta_{\alpha,+} - \delta_{\alpha,-}\right) - 2 \gamma \right) \tilde{\chi}_\alpha = \notag \\
- \frac{\im}{\hbar} \left[ \tilde{H}_\mathrm{S} , \tilde{\chi}_\alpha \right] + \gamma r_\alpha \tilde{\rho}_\mathrm{TT} \left( 0 \right) \, .
\label{eq:condchi}
\end{align}
Here, $\delta_{\alpha,+}$ and $\delta_{\alpha,-}$ denote the Kronecker symbol.
The operators $r_\alpha$ are defined as $r_0 = \ket{\up} \bra{\up} + \ket{\down} \bra{\down}$, $r_+ = \ket{\down} \bra{\up}$ and $r_- = \ket{\up} \bra{\down}$.

Eq.~\eqref{eq:condchi} can be solved by transforming into the eigenbasis of $\tilde{H}_{\rm S} = S D S^\dag$, where $D$ is diagonal.
We introduce $\tilde{\chi}_\alpha' = S^\dag \tilde{\chi}_\alpha S$ and consider the transformed Eq.~\eqref{eq:condchi} element-wise.
Rearranging for the elements of $\tilde{\chi}'_\alpha$, we obtain
\begin{align}
\tilde{\chi}'_\alpha = G_\alpha \circ \left( S^\dag \left( r_\alpha \tilde{\rho}_\mathrm{TT} \left( 0 \right) \right) S \right)
\label{eq:chiprime}
\end{align}
with a Schur product denoted by "$\circ$".
The elements of operator $G_\alpha$ are given by
\begin{align}
\left( G_\alpha \right)_{a,b} = \gamma \Big\{ -2 \gamma + \im \omega_{N} \left( \delta_{\alpha,+} - \delta_{\alpha,-} \right) \Big. \notag \\
\Big. + \im \left( \left( D \right)_{a,a} - \left( D \right)_{b,b} \right) \Big\}^{-1} \, .
\end{align}
Finally, the operators $\tilde{\chi}_\alpha$ result from transforming from the eigenbasis of $\tilde{H}_{\rm S}$ back into the original basis.

Altogether, this approach allows us to diagonalize $\tilde{H}_{\rm S}$ and prepare the three operators $G_\alpha$ before the simulation of a pulse sequence.
During the pulse sequence, the operator $\tilde{\rho}_\mathrm{TT} \left( 0 \right)$ after each pump pulse has to be inserted in Eq.~\eqref{eq:chiprime}.
The results for $\tilde{\chi}_\alpha'$ are transformed via $\tilde{\chi}_\alpha = S \tilde{\chi}_\alpha' S^\dag$ and then enter the time evolution of $\tilde{\rho}_{\rm S}$ in Eq.~\eqref{eq:timeevolrhos}.

\section{Revival amplitude of the electronic steady-state}
\label{app:elss}

\begin{figure}[b]
\begin{center}
\includegraphics[scale=1]{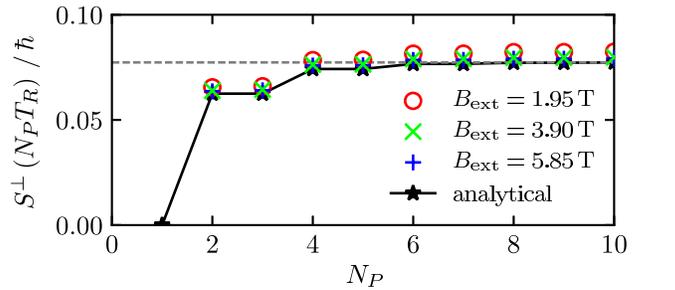}
\caption{(Color online) Evolution of the electron spin revival amplitude for small numbers $\NP$ of pump pulses, {i.\,e.}, without nuclear frequency focusing. Various colors show the amplitudes for different external magnetic fields $\Bext$. The black curve is calculated analytically from Eq.~\eqref{eq:elrevamp}. The analytic revival amplitude $\left| \SzTR_\infty \right| = \left| 1/2 - 1/\sqrt{3} \right| \approx 0.077$ in the limit $\NP \rightarrow \infty$ is indicated by a grey dashed horizontal line.}
\label{fig:elrevamp}
\end{center}
\end{figure}

Even before the nuclear spins are affected by the pump pulses, a revival amplitude appears due to a purely electronic steady-state \cite{BeugelingUhrigAnders2016}.
The evolution of this electronic revival can be understood by iteration of the pump pulse (cf. Eq.~\eqref{eq:sxpump}) and the evolution for the time $\TR$ (cf. Eq.~\eqref{eq:sxevol}).
Similar to the calculation of the revival amplitude in Eq.~\eqref{eq:sxtr}, we first consider a single nuclear configuration $K$ for a set $j$ of couplings.
The iteration of Eq.~\eqref{eq:sxpump} and Eq.~\eqref{eq:sxevol} yields
\begin{align}
\SzTR _{K,j}^{b} = - \sum_{i=1}^{\NP} \frac{\hbar}{2^{i+1}} \left\{  \cos \Big( \left( \omega_{e} + \omega_{K,j} \right) \TR \Big) \right\}^i
\label{eq:elrevampKj}
\end{align}
after $\NP$ pump pulses.
If we assume our external magnetic field to ensure an integer number of electron spin revolutions between two pump pulses, $\omega_{e} \TR$ is an integer multiple of $2 \pi$ and can be omitted in the cosine.
The full revival amplitude results from integrating over $\omega_{K,j}$ weighted by its Gaussian distribution as we do not consider nuclear focusing.
Since the width of the Gaussian distribution of $\omega_{K,j}$ is proportional to the inverse $T^*$, it is large compared to the periodicity of the cosine in Eq.~\eqref{eq:elrevampKj} that is determined by the inverse of $\TR$.
Hence, we substitute the integration of Eq.~\eqref{eq:elrevampKj} over $\omega_{K,j}$ by an integration over one period of the cosine ($\omega_{K,j} \in \left[ 0 ; 2\pi / \TR \right]$)
\begin{align}
\SzTR &= - \sum_{i=1}^{\NP} \frac{\hbar}{2^{i+2}} \int_0^{2\pi / \TR} \mathrm{d}\omega_{K,j}  \left(  \cos \left( \omega_{K,j} \TR \right) \right)^i \notag \\
&= - \sum_{i=1}^{\lfloor \NP /2 \rfloor} \frac{\hbar}{2^{4i+1}} \, \frac{(2i)!}{(i!)^{2}} \, .
\label{eq:elrevamp}
\end{align}
Since the integral over $\omega_{K,j}$ yields zero for odd $i$, we transform the index of summation $i \rightarrow i/2$ in the second line of Eq.~\eqref{eq:elrevamp}. From a physical point of view, the contributions to the revival amplitude from different Overhauser fields cancel each other for every second pulse.
Thus, the revival amplitude increases with $\NP$ in steps of two.

Note that we obtain
\begin{align}
\SyTR_{K,j}^{b} = &- \sum_{i=1}^{\NP} \frac{\hbar}{2^{i+1}} \left(  \cos \left( \left( \omega_{e} + \omega_{K,j} \right) \TR \right) \right)^{i-1} \notag \\
& \cdot \sin \left( \left( \omega_{e} + \omega_{K,j} \right) \TR \right)
\end{align}
for the spin component in $y$-direction.
Thus, the integration in analogy to Eq.~\eqref{eq:elrevamp} yields $\SyTR = 0$ and we can state $\SpTR = \left| \SzTR \right|$in the analytic calculation.

The limit $\NP \rightarrow \infty$ yields the final revival amplitude of the electronic steady-state $\SzTR_\infty = 1/2 - 1/\sqrt{3} \approx -0.077$.
In Fig.~\ref{fig:elrevamp}, the growth of revival amplitude up to the tenth pump pulse is illustrated.
The deviations of the numerical calculations (colored symbols) from Eq.~\eqref{eq:elrevamp} (black curve) are only minor and due to the finite number of nuclear spins ($N=6$).



\begin{thebibliography}{52}%
\makeatletter
\providecommand \@ifxundefined [1]{%
 \@ifx{#1\undefined}
}%
\providecommand \@ifnum [1]{%
 \ifnum #1\expandafter \@firstoftwo
 \else \expandafter \@secondoftwo
 \fi
}%
\providecommand \@ifx [1]{%
 \ifx #1\expandafter \@firstoftwo
 \else \expandafter \@secondoftwo
 \fi
}%
\providecommand \natexlab [1]{#1}%
\providecommand \enquote  [1]{``#1''}%
\providecommand \bibnamefont  [1]{#1}%
\providecommand \bibfnamefont [1]{#1}%
\providecommand \citenamefont [1]{#1}%
\providecommand \href@noop [0]{\@secondoftwo}%
\providecommand \href [0]{\begingroup \@sanitize@url \@href}%
\providecommand \@href[1]{\@@startlink{#1}\@@href}%
\providecommand \@@href[1]{\endgroup#1\@@endlink}%
\providecommand \@sanitize@url [0]{\catcode `\\12\catcode `\$12\catcode
  `\&12\catcode `\#12\catcode `\^12\catcode `\_12\catcode `\%12\relax}%
\providecommand \@@startlink[1]{}%
\providecommand \@@endlink[0]{}%
\providecommand \url  [0]{\begingroup\@sanitize@url \@url }%
\providecommand \@url [1]{\endgroup\@href {#1}{\urlprefix }}%
\providecommand \urlprefix  [0]{URL }%
\providecommand \Eprint [0]{\href }%
\providecommand \doibase [0]{http://dx.doi.org/}%
\providecommand \selectlanguage [0]{\@gobble}%
\providecommand \bibinfo  [0]{\@secondoftwo}%
\providecommand \bibfield  [0]{\@secondoftwo}%
\providecommand \translation [1]{[#1]}%
\providecommand \BibitemOpen [0]{}%
\providecommand \bibitemStop [0]{}%
\providecommand \bibitemNoStop [0]{.\EOS\space}%
\providecommand \EOS [0]{\spacefactor3000\relax}%
\providecommand \BibitemShut  [1]{\csname bibitem#1\endcsname}%
\let\auto@bib@innerbib\@empty
\bibitem [{\citenamefont {Greilich}\ \emph
  {et~al.}(2006{\natexlab{a}})\citenamefont {Greilich}, \citenamefont {Oulton},
  \citenamefont {Zhukov}, \citenamefont {Yugova}, \citenamefont {Yakovlev},
  \citenamefont {Bayer}, \citenamefont {Shabaev}, \citenamefont {Efros},
  \citenamefont {Merkulov}, \citenamefont {Stavarache}, \citenamefont
  {Reuter},\ and\ \citenamefont {Wieck}}]{PhysRevLett.96.227401}%
  \BibitemOpen
  \bibfield  {author} {\bibinfo {author} {\bibfnamefont {A.}~\bibnamefont
  {Greilich}}, \bibinfo {author} {\bibfnamefont {R.}~\bibnamefont {Oulton}},
  \bibinfo {author} {\bibfnamefont {E.~A.}\ \bibnamefont {Zhukov}}, \bibinfo
  {author} {\bibfnamefont {I.~A.}\ \bibnamefont {Yugova}}, \bibinfo {author}
  {\bibfnamefont {D.~R.}\ \bibnamefont {Yakovlev}}, \bibinfo {author}
  {\bibfnamefont {M.}~\bibnamefont {Bayer}}, \bibinfo {author} {\bibfnamefont
  {A.}~\bibnamefont {Shabaev}}, \bibinfo {author} {\bibfnamefont {{\relax
  Al}.~L.}\ \bibnamefont {Efros}}, \bibinfo {author} {\bibfnamefont {I.~A.}\
  \bibnamefont {Merkulov}}, \bibinfo {author} {\bibfnamefont {V.}~\bibnamefont
  {Stavarache}}, \bibinfo {author} {\bibfnamefont {D.}~\bibnamefont {Reuter}},
  \ and\ \bibinfo {author} {\bibfnamefont {A.}~\bibnamefont {Wieck}},\ }\href
  {\doibase 10.1103/PhysRevLett.96.227401} {\bibfield  {journal} {\bibinfo
  {journal} {Phys. Rev. Lett.}\ }\textbf {\bibinfo {volume} {96}},\ \bibinfo
  {pages} {227401} (\bibinfo {year} {2006}{\natexlab{a}})}\BibitemShut
  {NoStop}%
\bibitem [{\citenamefont {Greilich}\ \emph {et~al.}(2007)\citenamefont
  {Greilich}, \citenamefont {Shabaev}, \citenamefont {Yakovlev}, \citenamefont
  {Efros}, \citenamefont {Yugova}, \citenamefont {Reuter}, \citenamefont
  {Wieck},\ and\ \citenamefont {Bayer}}]{GreilichScience2007}%
  \BibitemOpen
  \bibfield  {author} {\bibinfo {author} {\bibfnamefont {A.}~\bibnamefont
  {Greilich}}, \bibinfo {author} {\bibfnamefont {A.}~\bibnamefont {Shabaev}},
  \bibinfo {author} {\bibfnamefont {D.~R.}\ \bibnamefont {Yakovlev}}, \bibinfo
  {author} {\bibfnamefont {{\relax Al}.~L.}\ \bibnamefont {Efros}}, \bibinfo
  {author} {\bibfnamefont {I.~A.}\ \bibnamefont {Yugova}}, \bibinfo {author}
  {\bibfnamefont {D.}~\bibnamefont {Reuter}}, \bibinfo {author} {\bibfnamefont
  {A.~D.}\ \bibnamefont {Wieck}}, \ and\ \bibinfo {author} {\bibfnamefont
  {M.}~\bibnamefont {Bayer}},\ }\href {\doibase 10.1126/science.1146850}
  {\bibfield  {journal} {\bibinfo  {journal} {Science}\ }\textbf {\bibinfo
  {volume} {317}},\ \bibinfo {pages} {1896} (\bibinfo {year}
  {2007})}\BibitemShut {NoStop}%
\bibitem [{\citenamefont {Merkulov}\ \emph {et~al.}(2002)\citenamefont
  {Merkulov}, \citenamefont {Efros},\ and\ \citenamefont
  {Rosen}}]{MerkulovEfrosRosen}%
  \BibitemOpen
  \bibfield  {author} {\bibinfo {author} {\bibfnamefont {I.~A.}\ \bibnamefont
  {Merkulov}}, \bibinfo {author} {\bibfnamefont {{\relax Al}.~L.}\ \bibnamefont
  {Efros}}, \ and\ \bibinfo {author} {\bibfnamefont {M.}~\bibnamefont
  {Rosen}},\ }\href {\doibase 10.1103/PhysRevB.65.205309} {\bibfield  {journal}
  {\bibinfo  {journal} {Phys. Rev. B}\ }\textbf {\bibinfo {volume} {65}},\
  \bibinfo {pages} {205309} (\bibinfo {year} {2002})}\BibitemShut {NoStop}%
\bibitem [{\citenamefont {Coish}\ and\ \citenamefont
  {Loss}(2004)}]{CoishLoss2004}%
  \BibitemOpen
  \bibfield  {author} {\bibinfo {author} {\bibfnamefont {W.~A.}\ \bibnamefont
  {Coish}}\ and\ \bibinfo {author} {\bibfnamefont {D.}~\bibnamefont {Loss}},\
  }\href {\doibase 10.1103/PhysRevB.70.195340} {\bibfield  {journal} {\bibinfo
  {journal} {Phys. Rev. B}\ }\textbf {\bibinfo {volume} {70}},\ \bibinfo
  {pages} {195340} (\bibinfo {year} {2004})}\BibitemShut {NoStop}%
\bibitem [{\citenamefont {Fischer}\ \emph {et~al.}(2008)\citenamefont
  {Fischer}, \citenamefont {Coish}, \citenamefont {Bulaev},\ and\ \citenamefont
  {Loss}}]{FischerLoss2008}%
  \BibitemOpen
  \bibfield  {author} {\bibinfo {author} {\bibfnamefont {J.}~\bibnamefont
  {Fischer}}, \bibinfo {author} {\bibfnamefont {W.~A.}\ \bibnamefont {Coish}},
  \bibinfo {author} {\bibfnamefont {D.~V.}\ \bibnamefont {Bulaev}}, \ and\
  \bibinfo {author} {\bibfnamefont {D.}~\bibnamefont {Loss}},\ }\href@noop {}
  {\bibfield  {journal} {\bibinfo  {journal} {Phys. Rev. B}\ }\textbf {\bibinfo
  {volume} {78}},\ \bibinfo {pages} {155329} (\bibinfo {year}
  {2008})}\BibitemShut {NoStop}%
\bibitem [{\citenamefont {Hanson}\ \emph {et~al.}(2007)\citenamefont {Hanson},
  \citenamefont {Kouwenhoven}, \citenamefont {Petta}, \citenamefont {Tarucha},\
  and\ \citenamefont {Vandersypen}}]{HansonSpinQdotsRMP2007}%
  \BibitemOpen
  \bibfield  {author} {\bibinfo {author} {\bibfnamefont {R.}~\bibnamefont
  {Hanson}}, \bibinfo {author} {\bibfnamefont {L.~P.}\ \bibnamefont
  {Kouwenhoven}}, \bibinfo {author} {\bibfnamefont {J.~R.}\ \bibnamefont
  {Petta}}, \bibinfo {author} {\bibfnamefont {S.}~\bibnamefont {Tarucha}}, \
  and\ \bibinfo {author} {\bibfnamefont {L.~M.~K.}\ \bibnamefont
  {Vandersypen}},\ }\href {\doibase 10.1103/RevModPhys.79.1217} {\bibfield
  {journal} {\bibinfo  {journal} {Rev. Mod. Phys.}\ }\textbf {\bibinfo {volume}
  {79}},\ \bibinfo {pages} {1217} (\bibinfo {year} {2007})}\BibitemShut
  {NoStop}%
\bibitem [{\citenamefont {(Editor)}(2008)}]{Dyakonov}%
  \BibitemOpen
  \bibfield  {author} {\bibinfo {author} {\bibfnamefont {M.~I.~Dyakonov.}\
  \bibnamefont {(Editor)}},\ }\href@noop {} {\emph {\bibinfo {title} {Spin
  Physics in Semiconductors}}},\ \bibinfo {edition} {1st}\ ed.\ (\bibinfo
  {publisher} {Springer-Verlag Berlin Heidelberg},\ \bibinfo {year}
  {2008})\BibitemShut {NoStop}%
\bibitem [{\citenamefont {Urbaszek}\ \emph {et~al.}(2013)\citenamefont
  {Urbaszek}, \citenamefont {Marie}, \citenamefont {Amand}, \citenamefont
  {Krebs}, \citenamefont {Voisin}, \citenamefont {Maletinsky}, \citenamefont
  {H\"ogele},\ and\ \citenamefont {Imamoglu}}]{Urbaszek2013}%
  \BibitemOpen
  \bibfield  {author} {\bibinfo {author} {\bibfnamefont {B.}~\bibnamefont
  {Urbaszek}}, \bibinfo {author} {\bibfnamefont {X.}~\bibnamefont {Marie}},
  \bibinfo {author} {\bibfnamefont {T.}~\bibnamefont {Amand}}, \bibinfo
  {author} {\bibfnamefont {O.}~\bibnamefont {Krebs}}, \bibinfo {author}
  {\bibfnamefont {P.}~\bibnamefont {Voisin}}, \bibinfo {author} {\bibfnamefont
  {P.}~\bibnamefont {Maletinsky}}, \bibinfo {author} {\bibfnamefont
  {A.}~\bibnamefont {H\"ogele}}, \ and\ \bibinfo {author} {\bibfnamefont
  {A.}~\bibnamefont {Imamoglu}},\ }\href {\doibase 10.1103/RevModPhys.85.79}
  {\bibfield  {journal} {\bibinfo  {journal} {Rev. Mod. Phys.}\ }\textbf
  {\bibinfo {volume} {85}},\ \bibinfo {pages} {79} (\bibinfo {year}
  {2013})}\BibitemShut {NoStop}%
\bibitem [{\citenamefont {Petrov}\ and\ \citenamefont
  {Yakovlev}(2012)}]{PetrovYakovlev2012}%
  \BibitemOpen
  \bibfield  {author} {\bibinfo {author} {\bibfnamefont {M.~Y.}\ \bibnamefont
  {Petrov}}\ and\ \bibinfo {author} {\bibfnamefont {S.~V.}\ \bibnamefont
  {Yakovlev}},\ }\href {\doibase 10.1134/S1063776112060131} {\bibfield
  {journal} {\bibinfo  {journal} {Journal of Experimental and Theoretical
  Physics}\ }\textbf {\bibinfo {volume} {115}},\ \bibinfo {pages} {326}
  (\bibinfo {year} {2012})}\BibitemShut {NoStop}%
\bibitem [{\citenamefont {J\"aschke}\ \emph {et~al.}(2017)\citenamefont
  {J\"aschke}, \citenamefont {Fischer}, \citenamefont {Evers}, \citenamefont
  {Belykh}, \citenamefont {Greilich}, \citenamefont {Bayer},\ and\
  \citenamefont {Anders}}]{Jaeschke2017}%
  \BibitemOpen
  \bibfield  {author} {\bibinfo {author} {\bibfnamefont {N.}~\bibnamefont
  {J\"aschke}}, \bibinfo {author} {\bibfnamefont {A.}~\bibnamefont {Fischer}},
  \bibinfo {author} {\bibfnamefont {E.}~\bibnamefont {Evers}}, \bibinfo
  {author} {\bibfnamefont {V.~V.}\ \bibnamefont {Belykh}}, \bibinfo {author}
  {\bibfnamefont {A.}~\bibnamefont {Greilich}}, \bibinfo {author}
  {\bibfnamefont {M.}~\bibnamefont {Bayer}}, \ and\ \bibinfo {author}
  {\bibfnamefont {F.~B.}\ \bibnamefont {Anders}},\ }\href {\doibase
  10.1103/PhysRevB.96.205419} {\bibfield  {journal} {\bibinfo  {journal} {Phys.
  Rev. B}\ }\textbf {\bibinfo {volume} {96}},\ \bibinfo {pages} {205419}
  (\bibinfo {year} {2017})}\BibitemShut {NoStop}%
\bibitem [{\citenamefont {Beugeling}\ \emph {et~al.}(2016)\citenamefont
  {Beugeling}, \citenamefont {Uhrig},\ and\ \citenamefont
  {Anders}}]{BeugelingUhrigAnders2016}%
  \BibitemOpen
  \bibfield  {author} {\bibinfo {author} {\bibfnamefont {W.}~\bibnamefont
  {Beugeling}}, \bibinfo {author} {\bibfnamefont {G.~S.}\ \bibnamefont
  {Uhrig}}, \ and\ \bibinfo {author} {\bibfnamefont {F.~B.}\ \bibnamefont
  {Anders}},\ }\href {\doibase 10.1103/PhysRevB.94.245308} {\bibfield
  {journal} {\bibinfo  {journal} {Phys. Rev. B}\ }\textbf {\bibinfo {volume}
  {94}},\ \bibinfo {pages} {245308} (\bibinfo {year} {2016})}\BibitemShut
  {NoStop}%
\bibitem [{\citenamefont {Beugeling}\ \emph {et~al.}(2017)\citenamefont
  {Beugeling}, \citenamefont {Uhrig},\ and\ \citenamefont
  {Anders}}]{BeugelingUhrigAnders2017}%
  \BibitemOpen
  \bibfield  {author} {\bibinfo {author} {\bibfnamefont {W.}~\bibnamefont
  {Beugeling}}, \bibinfo {author} {\bibfnamefont {G.~S.}\ \bibnamefont
  {Uhrig}}, \ and\ \bibinfo {author} {\bibfnamefont {F.~B.}\ \bibnamefont
  {Anders}},\ }\href {\doibase 10.1103/PhysRevB.96.115303} {\bibfield
  {journal} {\bibinfo  {journal} {Phys. Rev. B}\ }\textbf {\bibinfo {volume}
  {96}},\ \bibinfo {pages} {115303} (\bibinfo {year} {2017})}\BibitemShut
  {NoStop}%
\bibitem [{\citenamefont {Schering}\ \emph {et~al.}(2018)\citenamefont
  {Schering}, \citenamefont {H\"udepohl}, \citenamefont {Uhrig},\ and\
  \citenamefont {Fauseweh}}]{Schering2018}%
  \BibitemOpen
  \bibfield  {author} {\bibinfo {author} {\bibfnamefont {P.}~\bibnamefont
  {Schering}}, \bibinfo {author} {\bibfnamefont {J.}~\bibnamefont
  {H\"udepohl}}, \bibinfo {author} {\bibfnamefont {G.~S.}\ \bibnamefont
  {Uhrig}}, \ and\ \bibinfo {author} {\bibfnamefont {B.}~\bibnamefont
  {Fauseweh}},\ }\href {\doibase 10.1103/PhysRevB.98.024305} {\bibfield
  {journal} {\bibinfo  {journal} {Phys. Rev. B}\ }\textbf {\bibinfo {volume}
  {98}},\ \bibinfo {pages} {024305} (\bibinfo {year} {2018})}\BibitemShut
  {NoStop}%
\bibitem [{\citenamefont {{Gaudin, M.}}(1976)}]{Gaudin}%
  \BibitemOpen
  \bibfield  {author} {\bibinfo {author} {\bibnamefont {{Gaudin, M.}}},\ }\href
  {\doibase 10.1051/jphys:0197600370100108700} {\bibfield  {journal} {\bibinfo
  {journal} {J. Phys. France}\ }\textbf {\bibinfo {volume} {37}},\ \bibinfo
  {pages} {1087} (\bibinfo {year} {1976})}\BibitemShut {NoStop}%
\bibitem [{\citenamefont {Al-Hassanieh}\ \emph {et~al.}(2006)\citenamefont
  {Al-Hassanieh}, \citenamefont {Dobrovitski}, \citenamefont {Dagotto},\ and\
  \citenamefont {Harmon}}]{alhas06}%
  \BibitemOpen
  \bibfield  {author} {\bibinfo {author} {\bibfnamefont {K.~A.}\ \bibnamefont
  {Al-Hassanieh}}, \bibinfo {author} {\bibfnamefont {V.~V.}\ \bibnamefont
  {Dobrovitski}}, \bibinfo {author} {\bibfnamefont {E.}~\bibnamefont
  {Dagotto}}, \ and\ \bibinfo {author} {\bibfnamefont {B.~N.}\ \bibnamefont
  {Harmon}},\ }\href@noop {} {\bibfield  {journal} {\bibinfo  {journal} {Phys.
  Rev. Lett.}\ }\textbf {\bibinfo {volume} {97}},\ \bibinfo {pages} {037204}
  (\bibinfo {year} {2006})}\BibitemShut {NoStop}%
\bibitem [{\citenamefont {Hackmann}\ and\ \citenamefont
  {Anders}(2014)}]{HackmannAnders2014}%
  \BibitemOpen
  \bibfield  {author} {\bibinfo {author} {\bibfnamefont {J.}~\bibnamefont
  {Hackmann}}\ and\ \bibinfo {author} {\bibfnamefont {F.~B.}\ \bibnamefont
  {Anders}},\ }\href {\doibase 10.1103/PhysRevB.89.045317} {\bibfield
  {journal} {\bibinfo  {journal} {Phys. Rev. B}\ }\textbf {\bibinfo {volume}
  {89}},\ \bibinfo {pages} {045317} (\bibinfo {year} {2014})}\BibitemShut
  {NoStop}%
\bibitem [{\citenamefont {Fr\"ohling}\ and\ \citenamefont
  {Anders}(2017)}]{FroehlingAnders2017}%
  \BibitemOpen
  \bibfield  {author} {\bibinfo {author} {\bibfnamefont {N.}~\bibnamefont
  {Fr\"ohling}}\ and\ \bibinfo {author} {\bibfnamefont {F.~B.}\ \bibnamefont
  {Anders}},\ }\href {\doibase 10.1103/PhysRevB.96.045441} {\bibfield
  {journal} {\bibinfo  {journal} {Phys. Rev. B}\ }\textbf {\bibinfo {volume}
  {96}},\ \bibinfo {pages} {045441} (\bibinfo {year} {2017})}\BibitemShut
  {NoStop}%
\bibitem [{\citenamefont {Warburton}(2013)}]{Warburton2013}%
  \BibitemOpen
  \bibfield  {author} {\bibinfo {author} {\bibfnamefont {R.~J.}\ \bibnamefont
  {Warburton}},\ }\href {http://dx.doi.org/10.1038/nmat3585} {\bibfield
  {journal} {\bibinfo  {journal} {Nature Materials}\ }\textbf {\bibinfo
  {volume} {12}},\ \bibinfo {pages} {483} (\bibinfo {year} {2013})}\BibitemShut
  {NoStop}%
\bibitem [{\citenamefont {Chekhovich}\ \emph {et~al.}(2013)\citenamefont
  {Chekhovich}, \citenamefont {Makhonin}, \citenamefont {Tartakovskii},
  \citenamefont {Yacoby}, \citenamefont {Bluhm}, \citenamefont {Nowack},\ and\
  \citenamefont {Vandersypen}}]{Chekhovich2013}%
  \BibitemOpen
  \bibfield  {author} {\bibinfo {author} {\bibfnamefont {E.~A.}\ \bibnamefont
  {Chekhovich}}, \bibinfo {author} {\bibfnamefont {M.~N.}\ \bibnamefont
  {Makhonin}}, \bibinfo {author} {\bibfnamefont {A.~I.}\ \bibnamefont
  {Tartakovskii}}, \bibinfo {author} {\bibfnamefont {A.}~\bibnamefont
  {Yacoby}}, \bibinfo {author} {\bibfnamefont {H.}~\bibnamefont {Bluhm}},
  \bibinfo {author} {\bibfnamefont {K.~C.}\ \bibnamefont {Nowack}}, \ and\
  \bibinfo {author} {\bibfnamefont {L.~M.~K.}\ \bibnamefont {Vandersypen}},\
  }\href {http://dx.doi.org/10.1038/nmat3652} {\bibfield  {journal} {\bibinfo
  {journal} {Nature Materials}\ }\textbf {\bibinfo {volume} {12}},\ \bibinfo
  {pages} {494} (\bibinfo {year} {2013})}\BibitemShut {NoStop}%
\bibitem [{\citenamefont {Press}\ \emph {et~al.}(2010)\citenamefont {Press},
  \citenamefont {De~Greve}, \citenamefont {McMahon}, \citenamefont {Ladd},
  \citenamefont {Friess}, \citenamefont {Schneider}, \citenamefont {Kamp},
  \citenamefont {H\"ofling}, \citenamefont {Forchel},\ and\ \citenamefont
  {Yamamoto}}]{Press2010}%
  \BibitemOpen
  \bibfield  {author} {\bibinfo {author} {\bibfnamefont {D.}~\bibnamefont
  {Press}}, \bibinfo {author} {\bibfnamefont {K.}~\bibnamefont {De~Greve}},
  \bibinfo {author} {\bibfnamefont {P.~L.}\ \bibnamefont {McMahon}}, \bibinfo
  {author} {\bibfnamefont {T.~D.}\ \bibnamefont {Ladd}}, \bibinfo {author}
  {\bibfnamefont {B.}~\bibnamefont {Friess}}, \bibinfo {author} {\bibfnamefont
  {C.}~\bibnamefont {Schneider}}, \bibinfo {author} {\bibfnamefont
  {M.}~\bibnamefont {Kamp}}, \bibinfo {author} {\bibfnamefont {S.}~\bibnamefont
  {H\"ofling}}, \bibinfo {author} {\bibfnamefont {A.}~\bibnamefont {Forchel}},
  \ and\ \bibinfo {author} {\bibfnamefont {Y.}~\bibnamefont {Yamamoto}},\
  }\href {\doibase 10.1038/nphoton.2010.83} {\bibfield  {journal} {\bibinfo
  {journal} {Nat Photon}\ }\textbf {\bibinfo {volume} {4}},\ \bibinfo {pages}
  {367 } (\bibinfo {year} {2010})}\BibitemShut {NoStop}%
\bibitem [{\citenamefont {Bechtold}\ \emph {et~al.}(2016)\citenamefont
  {Bechtold}, \citenamefont {Li}, \citenamefont {M\"uller}, \citenamefont
  {Simmet}, \citenamefont {Ardelt}, \citenamefont {Finley},\ and\ \citenamefont
  {Sinitsyn}}]{Bechtold2016}%
  \BibitemOpen
  \bibfield  {author} {\bibinfo {author} {\bibfnamefont {A.}~\bibnamefont
  {Bechtold}}, \bibinfo {author} {\bibfnamefont {F.}~\bibnamefont {Li}},
  \bibinfo {author} {\bibfnamefont {K.}~\bibnamefont {M\"uller}}, \bibinfo
  {author} {\bibfnamefont {T.}~\bibnamefont {Simmet}}, \bibinfo {author}
  {\bibfnamefont {P.-L.}\ \bibnamefont {Ardelt}}, \bibinfo {author}
  {\bibfnamefont {J.~J.}\ \bibnamefont {Finley}}, \ and\ \bibinfo {author}
  {\bibfnamefont {N.~A.}\ \bibnamefont {Sinitsyn}},\ }\href {\doibase
  10.1103/PhysRevLett.117.027402} {\bibfield  {journal} {\bibinfo  {journal}
  {Phys. Rev. Lett.}\ }\textbf {\bibinfo {volume} {117}},\ \bibinfo {pages}
  {027402} (\bibinfo {year} {2016})}\BibitemShut {NoStop}%
\bibitem [{\citenamefont {Chekhovich}\ \emph {et~al.}(2015)\citenamefont
  {Chekhovich}, \citenamefont {Hopkinson}, \citenamefont {Skolnick},\ and\
  \citenamefont {Tartakovskii}}]{Chekhovich2015}%
  \BibitemOpen
  \bibfield  {author} {\bibinfo {author} {\bibfnamefont {E.}~\bibnamefont
  {Chekhovich}}, \bibinfo {author} {\bibfnamefont {M.}~\bibnamefont
  {Hopkinson}}, \bibinfo {author} {\bibfnamefont {M.}~\bibnamefont {Skolnick}},
  \ and\ \bibinfo {author} {\bibfnamefont {A.}~\bibnamefont {Tartakovskii}},\
  }\href {http://dx.doi.org/10.1038/ncomms7348} {\bibfield  {journal} {\bibinfo
   {journal} {Nature Communications}\ }\textbf {\bibinfo {volume} {6}},\
  \bibinfo {pages} {6348} (\bibinfo {year} {2015})}\BibitemShut {NoStop}%
\bibitem [{\citenamefont {Bluhm}\ \emph {et~al.}(2010)\citenamefont {Bluhm},
  \citenamefont {Foletti}, \citenamefont {Mahalu}, \citenamefont {Umansky},\
  and\ \citenamefont {Yacoby}}]{Bluhm2010}%
  \BibitemOpen
  \bibfield  {author} {\bibinfo {author} {\bibfnamefont {H.}~\bibnamefont
  {Bluhm}}, \bibinfo {author} {\bibfnamefont {S.}~\bibnamefont {Foletti}},
  \bibinfo {author} {\bibfnamefont {D.}~\bibnamefont {Mahalu}}, \bibinfo
  {author} {\bibfnamefont {V.}~\bibnamefont {Umansky}}, \ and\ \bibinfo
  {author} {\bibfnamefont {A.}~\bibnamefont {Yacoby}},\ }\href {\doibase
  10.1103/PhysRevLett.105.216803} {\bibfield  {journal} {\bibinfo  {journal}
  {Phys. Rev. Lett.}\ }\textbf {\bibinfo {volume} {105}},\ \bibinfo {pages}
  {216803} (\bibinfo {year} {2010})}\BibitemShut {NoStop}%
\bibitem [{\citenamefont {Varwig}\ \emph {et~al.}(2014)\citenamefont {Varwig},
  \citenamefont {Evers}, \citenamefont {Greilich}, \citenamefont {Yakovlev},
  \citenamefont {Reuter}, \citenamefont {Wieck},\ and\ \citenamefont
  {Bayer}}]{Varwig2014}%
  \BibitemOpen
  \bibfield  {author} {\bibinfo {author} {\bibfnamefont {S.}~\bibnamefont
  {Varwig}}, \bibinfo {author} {\bibfnamefont {E.}~\bibnamefont {Evers}},
  \bibinfo {author} {\bibfnamefont {A.}~\bibnamefont {Greilich}}, \bibinfo
  {author} {\bibfnamefont {D.~R.}\ \bibnamefont {Yakovlev}}, \bibinfo {author}
  {\bibfnamefont {D.}~\bibnamefont {Reuter}}, \bibinfo {author} {\bibfnamefont
  {A.~D.}\ \bibnamefont {Wieck}}, \ and\ \bibinfo {author} {\bibfnamefont
  {M.}~\bibnamefont {Bayer}},\ }\href {\doibase 10.1103/PhysRevB.90.121306}
  {\bibfield  {journal} {\bibinfo  {journal} {Phys. Rev. B}\ }\textbf {\bibinfo
  {volume} {90}},\ \bibinfo {pages} {121306} (\bibinfo {year}
  {2014})}\BibitemShut {NoStop}%
\bibitem [{\citenamefont {Huthmacher}\ \emph {et~al.}(2018)\citenamefont
  {Huthmacher}, \citenamefont {Stockill}, \citenamefont {Clarke}, \citenamefont
  {Hugues}, \citenamefont {Le~Gall},\ and\ \citenamefont
  {Atat\"ure}}]{Huthmacher2018}%
  \BibitemOpen
  \bibfield  {author} {\bibinfo {author} {\bibfnamefont {L.}~\bibnamefont
  {Huthmacher}}, \bibinfo {author} {\bibfnamefont {R.}~\bibnamefont
  {Stockill}}, \bibinfo {author} {\bibfnamefont {E.}~\bibnamefont {Clarke}},
  \bibinfo {author} {\bibfnamefont {M.}~\bibnamefont {Hugues}}, \bibinfo
  {author} {\bibfnamefont {C.}~\bibnamefont {Le~Gall}}, \ and\ \bibinfo
  {author} {\bibfnamefont {M.}~\bibnamefont {Atat\"ure}},\ }\href {\doibase
  10.1103/PhysRevB.97.241413} {\bibfield  {journal} {\bibinfo  {journal} {Phys.
  Rev. B}\ }\textbf {\bibinfo {volume} {97}},\ \bibinfo {pages} {241413}
  (\bibinfo {year} {2018})}\BibitemShut {NoStop}%
\bibitem [{\citenamefont {Issler}\ \emph {et~al.}(2010)\citenamefont {Issler},
  \citenamefont {Kessler}, \citenamefont {Giedke}, \citenamefont {Yelin},
  \citenamefont {Cirac}, \citenamefont {Lukin},\ and\ \citenamefont
  {Imamoglu}}]{Issler2010}%
  \BibitemOpen
  \bibfield  {author} {\bibinfo {author} {\bibfnamefont {M.}~\bibnamefont
  {Issler}}, \bibinfo {author} {\bibfnamefont {E.~M.}\ \bibnamefont {Kessler}},
  \bibinfo {author} {\bibfnamefont {G.}~\bibnamefont {Giedke}}, \bibinfo
  {author} {\bibfnamefont {S.}~\bibnamefont {Yelin}}, \bibinfo {author}
  {\bibfnamefont {I.}~\bibnamefont {Cirac}}, \bibinfo {author} {\bibfnamefont
  {M.~D.}\ \bibnamefont {Lukin}}, \ and\ \bibinfo {author} {\bibfnamefont
  {A.}~\bibnamefont {Imamoglu}},\ }\href {\doibase
  10.1103/PhysRevLett.105.267202} {\bibfield  {journal} {\bibinfo  {journal}
  {Phys. Rev. Lett.}\ }\textbf {\bibinfo {volume} {105}},\ \bibinfo {pages}
  {267202} (\bibinfo {year} {2010})}\BibitemShut {NoStop}%
\bibitem [{\citenamefont {Prechtel}\ \emph {et~al.}(2016)\citenamefont
  {Prechtel}, \citenamefont {Kuhlmann}, \citenamefont {Houel}, \citenamefont
  {Ludwig}, \citenamefont {Valentin}, \citenamefont {Wieck},\ and\
  \citenamefont {Warburton}}]{Prechtel2016}%
  \BibitemOpen
  \bibfield  {author} {\bibinfo {author} {\bibfnamefont {J.~H.}\ \bibnamefont
  {Prechtel}}, \bibinfo {author} {\bibfnamefont {A.~V.}\ \bibnamefont
  {Kuhlmann}}, \bibinfo {author} {\bibfnamefont {J.}~\bibnamefont {Houel}},
  \bibinfo {author} {\bibfnamefont {A.}~\bibnamefont {Ludwig}}, \bibinfo
  {author} {\bibfnamefont {S.~R.}\ \bibnamefont {Valentin}}, \bibinfo {author}
  {\bibfnamefont {A.~D.}\ \bibnamefont {Wieck}}, \ and\ \bibinfo {author}
  {\bibfnamefont {R.~J.}\ \bibnamefont {Warburton}},\ }\href
  {http://dx.doi.org/10.1038/nmat4704} {\bibfield  {journal} {\bibinfo
  {journal} {Nature Materials}\ }\textbf {\bibinfo {volume} {15}},\ \bibinfo
  {pages} {981} (\bibinfo {year} {2016})}\BibitemShut {NoStop}%
\bibitem [{\citenamefont {\'Ethier-Majcher}\ \emph {et~al.}(2017)\citenamefont
  {\'Ethier-Majcher}, \citenamefont {Gangloff}, \citenamefont {Stockill},
  \citenamefont {Clarke}, \citenamefont {Hugues}, \citenamefont {Le~Gall},\
  and\ \citenamefont {Atat\"ure}}]{PhysRevLett.119.130503}%
  \BibitemOpen
  \bibfield  {author} {\bibinfo {author} {\bibfnamefont {G.}~\bibnamefont
  {\'Ethier-Majcher}}, \bibinfo {author} {\bibfnamefont {D.}~\bibnamefont
  {Gangloff}}, \bibinfo {author} {\bibfnamefont {R.}~\bibnamefont {Stockill}},
  \bibinfo {author} {\bibfnamefont {E.}~\bibnamefont {Clarke}}, \bibinfo
  {author} {\bibfnamefont {M.}~\bibnamefont {Hugues}}, \bibinfo {author}
  {\bibfnamefont {C.}~\bibnamefont {Le~Gall}}, \ and\ \bibinfo {author}
  {\bibfnamefont {M.}~\bibnamefont {Atat\"ure}},\ }\href {\doibase
  10.1103/PhysRevLett.119.130503} {\bibfield  {journal} {\bibinfo  {journal}
  {Phys. Rev. Lett.}\ }\textbf {\bibinfo {volume} {119}},\ \bibinfo {pages}
  {130503} (\bibinfo {year} {2017})}\BibitemShut {NoStop}%
\bibitem [{\citenamefont {W\"ust}\ \emph {et~al.}(2016)\citenamefont {W\"ust},
  \citenamefont {Munsch}, \citenamefont {Maier}, \citenamefont {Kuhlmann},
  \citenamefont {Ludwig}, \citenamefont {Wieck}, \citenamefont {Loss},
  \citenamefont {Poggio},\ and\ \citenamefont {Warburton}}]{Wuest2016}%
  \BibitemOpen
  \bibfield  {author} {\bibinfo {author} {\bibfnamefont {G.}~\bibnamefont
  {W\"ust}}, \bibinfo {author} {\bibfnamefont {M.}~\bibnamefont {Munsch}},
  \bibinfo {author} {\bibfnamefont {F.}~\bibnamefont {Maier}}, \bibinfo
  {author} {\bibfnamefont {A.~V.}\ \bibnamefont {Kuhlmann}}, \bibinfo {author}
  {\bibfnamefont {A.}~\bibnamefont {Ludwig}}, \bibinfo {author} {\bibfnamefont
  {A.~D.}\ \bibnamefont {Wieck}}, \bibinfo {author} {\bibfnamefont
  {D.}~\bibnamefont {Loss}}, \bibinfo {author} {\bibfnamefont {M.}~\bibnamefont
  {Poggio}}, \ and\ \bibinfo {author} {\bibfnamefont {R.~J.}\ \bibnamefont
  {Warburton}},\ }\href {http://dx.doi.org/10.1038/nnano.2016.114} {\bibfield
  {journal} {\bibinfo  {journal} {Nature Nanotechnology}\ }\textbf {\bibinfo
  {volume} {11}},\ \bibinfo {pages} {885} (\bibinfo {year} {2016})}\BibitemShut
  {NoStop}%
\bibitem [{\citenamefont {Shabaev}\ \emph {et~al.}(2003)\citenamefont
  {Shabaev}, \citenamefont {Efros}, \citenamefont {Gammon},\ and\ \citenamefont
  {Merkulov}}]{Shabaev2003}%
  \BibitemOpen
  \bibfield  {author} {\bibinfo {author} {\bibfnamefont {A.}~\bibnamefont
  {Shabaev}}, \bibinfo {author} {\bibfnamefont {{\relax Al}.~L.}\ \bibnamefont
  {Efros}}, \bibinfo {author} {\bibfnamefont {D.}~\bibnamefont {Gammon}}, \
  and\ \bibinfo {author} {\bibfnamefont {I.~A.}\ \bibnamefont {Merkulov}},\
  }\href {\doibase 10.1103/PhysRevB.68.201305} {\bibfield  {journal} {\bibinfo
  {journal} {Phys. Rev. B}\ }\textbf {\bibinfo {volume} {68}},\ \bibinfo
  {pages} {201305} (\bibinfo {year} {2003})}\BibitemShut {NoStop}%
\bibitem [{\citenamefont {Petrov}\ \emph {et~al.}(2008)\citenamefont {Petrov},
  \citenamefont {Ignatiev}, \citenamefont {Poltavtsev}, \citenamefont
  {Greilich}, \citenamefont {Bauschulte}, \citenamefont {Yakovlev},\ and\
  \citenamefont {Bayer}}]{Petrov2008}%
  \BibitemOpen
  \bibfield  {author} {\bibinfo {author} {\bibfnamefont {M.~Y.}\ \bibnamefont
  {Petrov}}, \bibinfo {author} {\bibfnamefont {I.~V.}\ \bibnamefont
  {Ignatiev}}, \bibinfo {author} {\bibfnamefont {S.~V.}\ \bibnamefont
  {Poltavtsev}}, \bibinfo {author} {\bibfnamefont {A.}~\bibnamefont
  {Greilich}}, \bibinfo {author} {\bibfnamefont {A.}~\bibnamefont
  {Bauschulte}}, \bibinfo {author} {\bibfnamefont {D.~R.}\ \bibnamefont
  {Yakovlev}}, \ and\ \bibinfo {author} {\bibfnamefont {M.}~\bibnamefont
  {Bayer}},\ }\href {\doibase 10.1103/PhysRevB.78.045315} {\bibfield  {journal}
  {\bibinfo  {journal} {Phys. Rev. B}\ }\textbf {\bibinfo {volume} {78}},\
  \bibinfo {pages} {045315} (\bibinfo {year} {2008})}\BibitemShut {NoStop}%
\bibitem [{\citenamefont {Greilich}\ \emph
  {et~al.}(2006{\natexlab{b}})\citenamefont {Greilich}, \citenamefont
  {Yakovlev}, \citenamefont {Shabaev}, \citenamefont {Efros}, \citenamefont
  {Yugova}, \citenamefont {Oulton}, \citenamefont {Stavarache}, \citenamefont
  {Reuter}, \citenamefont {Wieck},\ and\ \citenamefont
  {Bayer}}]{GreilichScience2006}%
  \BibitemOpen
  \bibfield  {author} {\bibinfo {author} {\bibfnamefont {A.}~\bibnamefont
  {Greilich}}, \bibinfo {author} {\bibfnamefont {D.~R.}\ \bibnamefont
  {Yakovlev}}, \bibinfo {author} {\bibfnamefont {A.}~\bibnamefont {Shabaev}},
  \bibinfo {author} {\bibfnamefont {{\relax Al}.~L.}\ \bibnamefont {Efros}},
  \bibinfo {author} {\bibfnamefont {I.~A.}\ \bibnamefont {Yugova}}, \bibinfo
  {author} {\bibfnamefont {R.}~\bibnamefont {Oulton}}, \bibinfo {author}
  {\bibfnamefont {V.}~\bibnamefont {Stavarache}}, \bibinfo {author}
  {\bibfnamefont {D.}~\bibnamefont {Reuter}}, \bibinfo {author} {\bibfnamefont
  {A.}~\bibnamefont {Wieck}}, \ and\ \bibinfo {author} {\bibfnamefont
  {M.}~\bibnamefont {Bayer}},\ }\href {\doibase 10.1126/science.1128215}
  {\bibfield  {journal} {\bibinfo  {journal} {Science}\ }\textbf {\bibinfo
  {volume} {313}},\ \bibinfo {pages} {341} (\bibinfo {year}
  {2006}{\natexlab{b}})}\BibitemShut {NoStop}%
\bibitem [{\citenamefont {Greilich}\ \emph {et~al.}(2009)\citenamefont
  {Greilich}, \citenamefont {Yakovlev},\ and\ \citenamefont
  {Bayer}}]{GREILICH20091466}%
  \BibitemOpen
  \bibfield  {author} {\bibinfo {author} {\bibfnamefont {A.}~\bibnamefont
  {Greilich}}, \bibinfo {author} {\bibfnamefont {D.~R.}\ \bibnamefont
  {Yakovlev}}, \ and\ \bibinfo {author} {\bibfnamefont {M.}~\bibnamefont
  {Bayer}},\ }\href {\doibase https://doi.org/10.1016/j.ssc.2009.04.045}
  {\bibfield  {journal} {\bibinfo  {journal} {Solid State Communications}\
  }\textbf {\bibinfo {volume} {149}},\ \bibinfo {pages} {1466 } (\bibinfo
  {year} {2009})}\BibitemShut {NoStop}%
\bibitem [{\citenamefont {Coish}\ and\ \citenamefont
  {Baugh}(2009)}]{CoishBaugh2009}%
  \BibitemOpen
  \bibfield  {author} {\bibinfo {author} {\bibfnamefont {W.~A.}\ \bibnamefont
  {Coish}}\ and\ \bibinfo {author} {\bibfnamefont {J.}~\bibnamefont {Baugh}},\
  }\href {\doibase 10.1002/pssb.200945229} {\bibfield  {journal} {\bibinfo
  {journal} {physica status solidi (b)}\ }\textbf {\bibinfo {volume} {246}},\
  \bibinfo {pages} {2203} (\bibinfo {year} {2009})}\BibitemShut {NoStop}%
\bibitem [{\citenamefont {Sinitsyn}\ \emph {et~al.}(2012)\citenamefont
  {Sinitsyn}, \citenamefont {Li}, \citenamefont {Crooker}, \citenamefont
  {Saxena},\ and\ \citenamefont {Smith}}]{Sinitsyn2012}%
  \BibitemOpen
  \bibfield  {author} {\bibinfo {author} {\bibfnamefont {N.~A.}\ \bibnamefont
  {Sinitsyn}}, \bibinfo {author} {\bibfnamefont {Y.}~\bibnamefont {Li}},
  \bibinfo {author} {\bibfnamefont {S.~A.}\ \bibnamefont {Crooker}}, \bibinfo
  {author} {\bibfnamefont {A.}~\bibnamefont {Saxena}}, \ and\ \bibinfo {author}
  {\bibfnamefont {D.~L.}\ \bibnamefont {Smith}},\ }\href {\doibase
  10.1103/PhysRevLett.109.166605} {\bibfield  {journal} {\bibinfo  {journal}
  {Phys. Rev. Lett.}\ }\textbf {\bibinfo {volume} {109}},\ \bibinfo {pages}
  {166605} (\bibinfo {year} {2012})}\BibitemShut {NoStop}%
\bibitem [{\citenamefont {Bulutay}(2012)}]{Bulutay2012}%
  \BibitemOpen
  \bibfield  {author} {\bibinfo {author} {\bibfnamefont {C.}~\bibnamefont
  {Bulutay}},\ }\href {\doibase 10.1103/PhysRevB.85.115313} {\bibfield
  {journal} {\bibinfo  {journal} {Phys. Rev. B}\ }\textbf {\bibinfo {volume}
  {85}},\ \bibinfo {pages} {115313} (\bibinfo {year} {2012})}\BibitemShut
  {NoStop}%
\bibitem [{\citenamefont {Bechtold}\ \emph {et~al.}(2015)\citenamefont
  {Bechtold}, \citenamefont {Rauch}, \citenamefont {Simmet}, \citenamefont
  {Ardelt}, \citenamefont {Regler}, \citenamefont {M\"uller}, \citenamefont
  {Sinitsyn},\ and\ \citenamefont {Finley}}]{Bechtold2015}%
  \BibitemOpen
  \bibfield  {author} {\bibinfo {author} {\bibfnamefont {A.}~\bibnamefont
  {Bechtold}}, \bibinfo {author} {\bibfnamefont {A.}~\bibnamefont {Rauch}},
  \bibinfo {author} {\bibfnamefont {T.}~\bibnamefont {Simmet}}, \bibinfo
  {author} {\bibfnamefont {P.-L.}\ \bibnamefont {Ardelt}}, \bibinfo {author}
  {\bibfnamefont {A.}~\bibnamefont {Regler}}, \bibinfo {author} {\bibfnamefont
  {K.}~\bibnamefont {M\"uller}}, \bibinfo {author} {\bibfnamefont {N.~A.}\
  \bibnamefont {Sinitsyn}}, \ and\ \bibinfo {author} {\bibfnamefont {J.~J.}\
  \bibnamefont {Finley}},\ }\href {\doibase 10.1038/nphys3470} {\bibfield
  {journal} {\bibinfo  {journal} {Nature Phys.}\ }\textbf {\bibinfo {volume}
  {11}},\ \bibinfo {pages} {1005} (\bibinfo {year} {2015})}\BibitemShut
  {NoStop}%
\bibitem [{\citenamefont {Hackmann}\ \emph {et~al.}(2015)\citenamefont
  {Hackmann}, \citenamefont {Glasenapp}, \citenamefont {Greilich},
  \citenamefont {Bayer},\ and\ \citenamefont {Anders}}]{hackmannPRL2015}%
  \BibitemOpen
  \bibfield  {author} {\bibinfo {author} {\bibfnamefont {J.}~\bibnamefont
  {Hackmann}}, \bibinfo {author} {\bibfnamefont {P.}~\bibnamefont {Glasenapp}},
  \bibinfo {author} {\bibfnamefont {A.}~\bibnamefont {Greilich}}, \bibinfo
  {author} {\bibfnamefont {M.}~\bibnamefont {Bayer}}, \ and\ \bibinfo {author}
  {\bibfnamefont {F.~B.}\ \bibnamefont {Anders}},\ }\href {\doibase
  10.1103/PhysRevLett.115.207401} {\bibfield  {journal} {\bibinfo  {journal}
  {Phys. Rev. Lett.}\ }\textbf {\bibinfo {volume} {115}},\ \bibinfo {pages}
  {207401} (\bibinfo {year} {2015})}\BibitemShut {NoStop}%
\bibitem [{\citenamefont {Glasenapp}\ \emph {et~al.}(2016)\citenamefont
  {Glasenapp}, \citenamefont {Smirnov}, \citenamefont {Greilich}, \citenamefont
  {Hackmann}, \citenamefont {Glazov}, \citenamefont {Anders},\ and\
  \citenamefont {Bayer}}]{Glasenapp2016}%
  \BibitemOpen
  \bibfield  {author} {\bibinfo {author} {\bibfnamefont {P.}~\bibnamefont
  {Glasenapp}}, \bibinfo {author} {\bibfnamefont {D.~S.}\ \bibnamefont
  {Smirnov}}, \bibinfo {author} {\bibfnamefont {A.}~\bibnamefont {Greilich}},
  \bibinfo {author} {\bibfnamefont {J.}~\bibnamefont {Hackmann}}, \bibinfo
  {author} {\bibfnamefont {M.~M.}\ \bibnamefont {Glazov}}, \bibinfo {author}
  {\bibfnamefont {F.~B.}\ \bibnamefont {Anders}}, \ and\ \bibinfo {author}
  {\bibfnamefont {M.}~\bibnamefont {Bayer}},\ }\href {\doibase
  10.1103/PhysRevB.93.205429} {\bibfield  {journal} {\bibinfo  {journal} {Phys.
  Rev. B}\ }\textbf {\bibinfo {volume} {93}},\ \bibinfo {pages} {205429}
  (\bibinfo {year} {2016})}\BibitemShut {NoStop}%
\bibitem [{\citenamefont {Faribault}\ and\ \citenamefont
  {Schuricht}(2013{\natexlab{a}})}]{FaribautSchuricht2013a}%
  \BibitemOpen
  \bibfield  {author} {\bibinfo {author} {\bibfnamefont {A.}~\bibnamefont
  {Faribault}}\ and\ \bibinfo {author} {\bibfnamefont {D.}~\bibnamefont
  {Schuricht}},\ }\href {\doibase 10.1103/PhysRevLett.110.040405} {\bibfield
  {journal} {\bibinfo  {journal} {Phys. Rev. Lett.}\ }\textbf {\bibinfo
  {volume} {110}},\ \bibinfo {pages} {040405} (\bibinfo {year}
  {2013}{\natexlab{a}})}\BibitemShut {NoStop}%
\bibitem [{\citenamefont {Faribault}\ and\ \citenamefont
  {Schuricht}(2013{\natexlab{b}})}]{FaribautSchuricht2013b}%
  \BibitemOpen
  \bibfield  {author} {\bibinfo {author} {\bibfnamefont {A.}~\bibnamefont
  {Faribault}}\ and\ \bibinfo {author} {\bibfnamefont {D.}~\bibnamefont
  {Schuricht}},\ }\href {\doibase 10.1103/PhysRevB.88.085323} {\bibfield
  {journal} {\bibinfo  {journal} {Phys. Rev. B}\ }\textbf {\bibinfo {volume}
  {88}},\ \bibinfo {pages} {085323} (\bibinfo {year}
  {2013}{\natexlab{b}})}\BibitemShut {NoStop}%
\bibitem [{\citenamefont {Testelin}\ \emph {et~al.}(2009)\citenamefont
  {Testelin}, \citenamefont {Bernardot}, \citenamefont {Eble},\ and\
  \citenamefont {Chamarro}}]{Testelin2009}%
  \BibitemOpen
  \bibfield  {author} {\bibinfo {author} {\bibfnamefont {C.}~\bibnamefont
  {Testelin}}, \bibinfo {author} {\bibfnamefont {F.}~\bibnamefont {Bernardot}},
  \bibinfo {author} {\bibfnamefont {B.}~\bibnamefont {Eble}}, \ and\ \bibinfo
  {author} {\bibfnamefont {M.}~\bibnamefont {Chamarro}},\ }\href {\doibase
  10.1103/PhysRevB.79.195440} {\bibfield  {journal} {\bibinfo  {journal} {Phys.
  Rev. B}\ }\textbf {\bibinfo {volume} {79}},\ \bibinfo {pages} {195440}
  (\bibinfo {year} {2009})}\BibitemShut {NoStop}%
\bibitem [{\citenamefont {Walchli}\ \emph {et~al.}(1953)\citenamefont
  {Walchli}, \citenamefont {Commission}, \citenamefont {Laboratory},
  \citenamefont {Carbide}, \citenamefont {Company}, \citenamefont {Carbide},\
  and\ \citenamefont {Corporation}}]{Walchli}%
  \BibitemOpen
  \bibfield  {author} {\bibinfo {author} {\bibfnamefont {H.}~\bibnamefont
  {Walchli}}, \bibinfo {author} {\bibfnamefont {U.~A.~E.}\ \bibnamefont
  {Commission}}, \bibinfo {author} {\bibfnamefont {O.~R.~N.}\ \bibnamefont
  {Laboratory}}, \bibinfo {author} {\bibnamefont {Carbide}}, \bibinfo {author}
  {\bibfnamefont {C.~C.}\ \bibnamefont {Company}}, \bibinfo {author}
  {\bibfnamefont {U.}~\bibnamefont {Carbide}}, \ and\ \bibinfo {author}
  {\bibfnamefont {C.}~\bibnamefont {Corporation}},\ }\href@noop {} {\emph
  {\bibinfo {title} {A table of nuclear moment data}}}\ (\bibinfo  {publisher}
  {Oak Ridge National Laboratory},\ \bibinfo {year} {1953})\BibitemShut
  {NoStop}%
\bibitem [{\citenamefont {Stone}(2005)}]{Stone}%
  \BibitemOpen
  \bibfield  {author} {\bibinfo {author} {\bibfnamefont {N.}~\bibnamefont
  {Stone}},\ }\href {\doibase https://doi.org/10.1016/j.adt.2005.04.001}
  {\bibfield  {journal} {\bibinfo  {journal} {Atomic Data and Nuclear Data
  Tables}\ }\textbf {\bibinfo {volume} {90}},\ \bibinfo {pages} {75 } (\bibinfo
  {year} {2005})}\BibitemShut {NoStop}%
\bibitem [{\citenamefont {Yugova}\ \emph {et~al.}(2012)\citenamefont {Yugova},
  \citenamefont {Glazov}, \citenamefont {Yakovlev}, \citenamefont {Sokolova},\
  and\ \citenamefont {Bayer}}]{PhysRevB.85.125304}%
  \BibitemOpen
  \bibfield  {author} {\bibinfo {author} {\bibfnamefont {I.~A.}\ \bibnamefont
  {Yugova}}, \bibinfo {author} {\bibfnamefont {M.~M.}\ \bibnamefont {Glazov}},
  \bibinfo {author} {\bibfnamefont {D.~R.}\ \bibnamefont {Yakovlev}}, \bibinfo
  {author} {\bibfnamefont {A.~A.}\ \bibnamefont {Sokolova}}, \ and\ \bibinfo
  {author} {\bibfnamefont {M.}~\bibnamefont {Bayer}},\ }\href {\doibase
  10.1103/PhysRevB.85.125304} {\bibfield  {journal} {\bibinfo  {journal} {Phys.
  Rev. B}\ }\textbf {\bibinfo {volume} {85}},\ \bibinfo {pages} {125304}
  (\bibinfo {year} {2012})}\BibitemShut {NoStop}%
\bibitem [{\citenamefont {Carmichael}(1999)}]{Carmichael}%
  \BibitemOpen
  \bibfield  {author} {\bibinfo {author} {\bibfnamefont {H.~J.}\ \bibnamefont
  {Carmichael}},\ }\href@noop {} {\emph {\bibinfo {title} {Statistical Methods
  in Quantum Optics 1}}}\ (\bibinfo  {publisher} {Springer Verlag},\ \bibinfo
  {address} {Berlin Heidelberg},\ \bibinfo {year} {1999})\BibitemShut {NoStop}%
\bibitem [{\citenamefont {Erlingsson}\ and\ \citenamefont
  {Nazarov}(2004)}]{erlin04}%
  \BibitemOpen
  \bibfield  {author} {\bibinfo {author} {\bibfnamefont {S.~I.}\ \bibnamefont
  {Erlingsson}}\ and\ \bibinfo {author} {\bibfnamefont {Y.~V.}\ \bibnamefont
  {Nazarov}},\ }\href@noop {} {\bibfield  {journal} {\bibinfo  {journal} {Phys.
  Rev. B}\ }\textbf {\bibinfo {volume} {70}},\ \bibinfo {pages} {205327}
  (\bibinfo {year} {2004})}\BibitemShut {NoStop}%
\bibitem [{\citenamefont {Chen}\ \emph {et~al.}(2007)\citenamefont {Chen},
  \citenamefont {Bergman},\ and\ \citenamefont {Balents}}]{Chen2007}%
  \BibitemOpen
  \bibfield  {author} {\bibinfo {author} {\bibfnamefont {G.}~\bibnamefont
  {Chen}}, \bibinfo {author} {\bibfnamefont {D.~L.}\ \bibnamefont {Bergman}}, \
  and\ \bibinfo {author} {\bibfnamefont {L.}~\bibnamefont {Balents}},\ }\href
  {\doibase 10.1103/PhysRevB.76.045312} {\bibfield  {journal} {\bibinfo
  {journal} {Phys. Rev. B}\ }\textbf {\bibinfo {volume} {76}},\ \bibinfo
  {pages} {045312} (\bibinfo {year} {2007})}\BibitemShut {NoStop}%
\bibitem [{\citenamefont {Polkovnikov}(2010)}]{polko10}%
  \BibitemOpen
  \bibfield  {author} {\bibinfo {author} {\bibfnamefont {A.}~\bibnamefont
  {Polkovnikov}},\ }\href {\doibase 10.1016/j.aop.2010.02.006} {\bibfield
  {journal} {\bibinfo  {journal} {Ann. of Phys.}\ }\textbf {\bibinfo {volume}
  {325}},\ \bibinfo {pages} {1790} (\bibinfo {year} {2010})}\BibitemShut
  {NoStop}%
\bibitem [{\citenamefont {Stanek}\ \emph {et~al.}(2014)\citenamefont {Stanek},
  \citenamefont {Raas},\ and\ \citenamefont {Uhrig}}]{Stanek2014}%
  \BibitemOpen
  \bibfield  {author} {\bibinfo {author} {\bibfnamefont {D.}~\bibnamefont
  {Stanek}}, \bibinfo {author} {\bibfnamefont {C.}~\bibnamefont {Raas}}, \ and\
  \bibinfo {author} {\bibfnamefont {G.~S.}\ \bibnamefont {Uhrig}},\ }\href
  {\doibase 10.1103/PhysRevB.90.064301} {\bibfield  {journal} {\bibinfo
  {journal} {Phys. Rev. B}\ }\textbf {\bibinfo {volume} {90}},\ \bibinfo
  {pages} {064301} (\bibinfo {year} {2014})}\BibitemShut {NoStop}%
\bibitem [{\citenamefont {Fauseweh}\ \emph {et~al.}(2017)\citenamefont
  {Fauseweh}, \citenamefont {Schering}, \citenamefont {H\"udepohl},\ and\
  \citenamefont {Uhrig}}]{Fauseweh2017}%
  \BibitemOpen
  \bibfield  {author} {\bibinfo {author} {\bibfnamefont {B.}~\bibnamefont
  {Fauseweh}}, \bibinfo {author} {\bibfnamefont {P.}~\bibnamefont {Schering}},
  \bibinfo {author} {\bibfnamefont {J.}~\bibnamefont {H\"udepohl}}, \ and\
  \bibinfo {author} {\bibfnamefont {G.~S.}\ \bibnamefont {Uhrig}},\ }\href
  {\doibase 10.1103/PhysRevB.96.054415} {\bibfield  {journal} {\bibinfo
  {journal} {Phys. Rev. B}\ }\textbf {\bibinfo {volume} {96}},\ \bibinfo
  {pages} {054415} (\bibinfo {year} {2017})}\BibitemShut {NoStop}%
\bibitem [{\citenamefont {R\"ohrig}\ \emph {et~al.}(2018)\citenamefont
  {R\"ohrig}, \citenamefont {Schering}, \citenamefont {Gravert}, \citenamefont
  {Fauseweh},\ and\ \citenamefont {Uhrig}}]{Roehrig2018}%
  \BibitemOpen
  \bibfield  {author} {\bibinfo {author} {\bibfnamefont {R.}~\bibnamefont
  {R\"ohrig}}, \bibinfo {author} {\bibfnamefont {P.}~\bibnamefont {Schering}},
  \bibinfo {author} {\bibfnamefont {L.~B.}\ \bibnamefont {Gravert}}, \bibinfo
  {author} {\bibfnamefont {B.}~\bibnamefont {Fauseweh}}, \ and\ \bibinfo
  {author} {\bibfnamefont {G.~S.}\ \bibnamefont {Uhrig}},\ }\href {\doibase
  10.1103/PhysRevB.97.165431} {\bibfield  {journal} {\bibinfo  {journal} {Phys.
  Rev. B}\ }\textbf {\bibinfo {volume} {97}},\ \bibinfo {pages} {165431}
  (\bibinfo {year} {2018})}\BibitemShut {NoStop}%
\end{thebibliography}

%

\end{document}